\documentclass{article}
\usepackage[top=1in, bottom=1in, left=1in, right=1in]{geometry}
\usepackage{amsmath}
\usepackage{graphicx}
\usepackage[colorlinks=true, allcolors=gray]{hyperref}
\usepackage{palatino} 
\usepackage[numbers,sort&compress]{natbib}
\usepackage{bm}
\usepackage{float}
\usepackage{xspace}
\usepackage{xcolor}
\usepackage{caption}
\usepackage[acronym]{glossaries}
\usepackage[bottom]{footmisc} 

\graphicspath{{./}{./figures}{./figures_si}}

\newcommand*{\fref}[1]{Fig.~\ref{#1}}
\newcommand*{\tref}[1]{Table~\ref{#1}}
\newcommand*{\eref}[1]{Eq.~\eqref{#1}}
\newcommand*{\sref}[1]{Section~\ref{#1}}

\newcommand*{\onlinecite}[1]{Ref.~\citenum{#1}}

\newcommand*{\net}{MatTen\xspace}
\newacronym{si}{ESI}{Electronic Supplementary Information}

\title{An Equivariant Graph Neural Network for the Elasticity Tensors of All Seven Crystal Systems}

\author{
\normalsize{Mingjian Wen$^1$\footnote{Email: mjwen@uh.edu},
Matthew K.\ Horton$^{2,3}$,
Jason M.\ Munro$^2$,
Patrick Huck$^4$,
Kristin A.\ Persson$^{5,6}$
}\\
\footnotesize{$^1$ Chemical and Biomolecular Engineering, University of Houston, Houston, 77204, TX, USA} \\
\footnotesize{$^2$ Materials Sciences Division, Lawrence Berkeley National Laboratory, Berkeley, 94720, CA, USA} \\
\footnotesize{$^3$ Microsoft Research, Redmond, 98052, WA, USA} \\
\footnotesize{$^4$  Energy Technologies Area, Lawrence Berkeley National Laboratory, Berkeley, 94720, CA, USA } \\
\footnotesize{$^5$  Molecular Foundry, Lawrence Berkeley National Laboratory, Berkeley, 94720, CA, USA} \\
\footnotesize{$^6$ Department of Materials Science and Engineering, University of California, Berkeley, Berkeley, 94720, CA, USA}
\vspace{-8mm}
}

\date{}

\begin{document}
\maketitle

\begin{abstract}

    The elasticity tensor is a fundamental material property that describes the elastic response of a material to external force.
    The availability of full elasticity tensors for inorganic crystalline compounds, however, is limited due to experimental and computational challenges.
    Here, we report the materials tensor (\net) model for rapid and accurate estimation of the full fourth-rank elasticity tensors of crystals.
    Based on equivariant graph neural networks, \net satisfies the two essential requirements for elasticity tensors: independence of the frame of reference and preservation of material symmetry.
    Consequently, it provides a unified treatment of elasticity tensors for all seven crystal systems across diverse chemical spaces, without the need to deal with each separately.
    \net was trained on a dataset of first-principles elasticity tensors garnered by the Materials Project over the past several years (we are releasing the data herein) and has broad applications in predicting the isotropic elastic properties of polycrystalline materials, examining the anisotropic behavior of single crystals, and discovering new materials with exceptional mechanical properties.
    Using \net, we have discovered a hundred new crystals with extremely large maximum directional Young's modulus and eleven polymorphs of elemental cubic metals with unconventional spatial orientation of Young's modulus.

\end{abstract}

\section{Introduction}
\label{sec:intro}

All materials exhibit elastic behavior under small external loads and return to their original shape upon the release of these loads \cite{hetnarski2016mathematical}.
The elasticity tensor provides a fundamental and complete description of a material's response to such loads.
It offers a lens to examine the inherent strength of the bonding in a material and enables the understanding, analyzing, and designing of many macroscopic physical properties of materials.
Commonly employed mechanical properties (for instance, Young's modulus and Poisson's ratio), thermal properties (for instance, thermal conductivity) and acoustic properties (for instance, sound velocity) can be mathematically derived from the elasticity tensor.
These properties have long been leveraged, for example, by materials scientists to search for ultrahard materials \cite{kaner2005designing,mansouri2018machine} and by geophysicists to interpret seismic data \cite{anderson1968some,karki2001high}.
More recently, the anisotropic elastic behavior of inorganic solid electrolytes has been found to play a decisive role in determining the stability of electrodeposition at the interfaces in solid-state batteries \cite{monroe2004effect,ahmad2017stability}.
Moreover, in solid-state synthesis, one would want to utilize the elasticity tensor to determine the local stability of a material, so as to avoid unsuccessful synthesis of materials that spontaneously transform into different structures \cite{mouhat2014necessary,tolborg2022free}.

In spite of the importance, elasticity tensor data from experimental measurement is limited to a small number of materials.
For example, for inorganic crystalline compounds, experimental data is only on the order of a few hundred, considering entries both tabulated in handbooks and scattered in individual papers \cite{de2015charting}.
The major difficulty lies in preparing large enough single crystals for accurate experimental measurement using current techniques such as resonant acoustic spectroscopy \cite{du2017facile}.
In the past decade, efficient and reliable electronic structure calculation methods such as density functional theory (DFT) \cite{lejaeghere2016reproducibility} with automation frameworks \cite{ong2013python, jain2015fireworks, mathew2017atomate} have enabled high-throughput computational investigation of materials.
Using this approach in the Materials Project \cite{jain2013commentary}, we produced an initial dataset of elasticity tensors for 1181 crystals in 2015 \cite{de2015charting}, which has been expanded over time to 10276, which we now release as a new dataset with this work.
Nevertheless, this only accounts for 6.6\% of the more than 154000 crystals in the Materials Project, let alone the even greater number of crystals recorded in the Inorganic Crystal Structure Database (ICSD) \cite{hellenbrandt2004inorganic} and predicted by universal interatomic potentials \cite{chen2022universal}.

It is therefore no surprise that machine learning (ML) has gathered substantial interest as a means to develop efficient surrogate models for the prediction of elastic properties.
In a nutshell, state-of-the-art ML models for elastic properties encode compositional information \cite{wang2021compositionally,dunn2020benchmarking,de2021materials} and/or structural information \cite{dunn2020benchmarking,de2021materials,chen2019graph,choudhary2021atomistic} in a material as feature vectors and then map them to a target using some regression algorithms.
This approach is adopted in many existing works for learning elastic properties of, e.g., alloys \cite{mukhamedov2021machine,vazquez2022efficient,linton2023machine, pasini2023graph} and polycrystals \cite{karimi2023prediction, hestroffer2023graph}.
They are, however, limited to derived scalar elastic properties such as bulk modulus and shear modulus, and separate models are built for each derived property.
Ideally, one would hope to predict the full elasticity tensor, from which all other elastic properties can be derived.
Along this direction, there have been attempts to predict individual tensor components \cite{ahmad2018machine, revi2021machine}.
These models are great first steps, but essentially they still predict separate scalars in a specific coordinate system and thus unavoidably violate the transformation rules of tensors.

Geometric machine learning \cite{bronstein2021geometric} such as equivariant graph neural networks (GNNs) \cite{thomas2018tensor,satorras2021n,batzner2022e3,takamoto2022towards,equiformer_v2} and equivariant kernel methods \cite{grisafi2018symmetry, veit2020predicting} can directly operate in the space of high-rank tensors and adhere to their transformation rules.
The main use case is still for scalar molecular and materials properties, but a couple of works have explored the application in predicting tensorial properties such as the molecular dipole moment \cite{pmlr-v139-schutt21a,veit2020predicting}, magnetic moment \cite{li2023deep}, and dielectric response \cite{grisafi2018symmetry}.
Other applications that do not directly predict final tensorial targets have also successfully taken advantage of internal tensorial representations to learn scalar fields such as molecular electron density \cite{unke2021se, rackers2023recipe}.
Although these efforts focus on scalars or low-rank tensors, they demonstrate the viability of direct machine learning of the full fourth-rank elasticity tensor.

In this work, we develop the Materials Tensor (\net) model for a rapid and accurate estimate of the fourth-rank elasticity tensors of inorganic compounds.
Our model, based on equivariant GNNs, takes a crystal structure as input and outputs its full elasticity tensor with all symmetry-related transformation rules automatically satisfied.
Other elastic properties such as bulk modulus and shear modulus can then be derived from the full elasticity tensor.
The model satisfies two essential symmetry requirements for elasticity tensors:
\emph{independence of the frame of reference}, meaning that the choice of a specific coordinate system does not affect the model output, and \emph{preservation of material symmetry}, meaning that the symmetry in a crystal is captured and reflected in the output elasticity tensor.
The capabilities of \net are demonstrated via the study of both isotropic and anisotropic elastic properties.
Using MatTen, we screened the Materials Project database for the identification of materials with a large maximum directional Young's modulus.
On average, the values of the newly found materials are more than three times larger than existing data, as verified by first-principles calculations.
In addition, we have identified 11 unconventional polymorphs of elemental cubic metals whose maximum directional Young's moduli are in the $\langle 100 \rangle$ directions.

\section{Results and Discussion}
\label{sec:results}

\subsection{Symmetry and irreducible representation of the elasticity tensor}
\label{sec:et:sym}

\begin{figure}[tbh!]
    \centering
    \includegraphics[width=0.65\columnwidth]{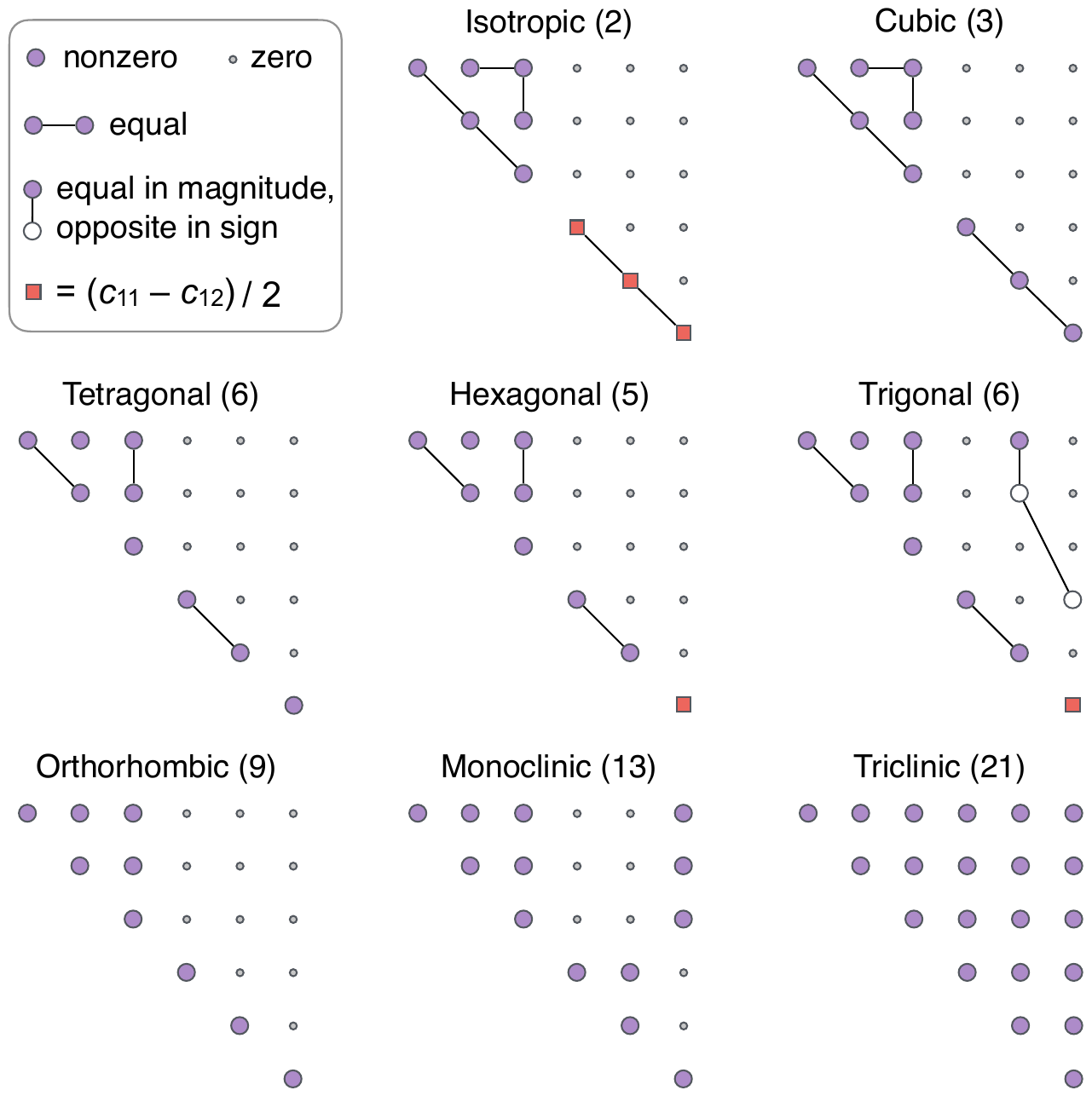}
    \caption{Symmetry classes and independent components of the elasticity tensor.
        The value in the parentheses after the name indicates the number of independent components.
        All matrices are symmetric about the main diagonal, with the lower triangular part omitted in the depiction.
        For single crystals considered in this work, the isotropic case does not apply.
        See \onlinecite{tadmor2012continuum} for a detailed treatment of the classification.
    }
    \label{fig:tensor:components}
\end{figure}

The elasticity tensor $\bm C$ is a fourth-rank tensor that fully characterizes the elastic behavior of a material.
Given that it is the second derivative of the total elastic energy with respect to the strain tensor and that the strain tensor is symmetric \cite{fedorov1968theory,nye1985physical}, the elasticity tensor possesses major symmetry $C_{ijkl} = C_{klij}$ and minor symmetry $C_{ijkl}= C_{jikl} = C_{ijlk}$ (in indicial notation, where $i,j,k,l \in \{1, 2, 3\}$).
Consequently, only 21 of the 81 components of $\bm C$ are independent.
It is convenient to write the elasticity tensor in a contracted matrix $\bm c$ ($c_{\alpha\beta}$, where $\alpha,\beta \in \{1,2,\dots,6\}$) with a pairs of indices $ij$ in the tensor notation replaced with a single index $\alpha$ in the matrix notation:
$11\rightarrow1$;
$22\rightarrow2$;
$33\rightarrow3$;
$23, 32\rightarrow4$;
$13, 31\rightarrow5$;
and
$12, 21\rightarrow6$.
This Voigt matrix \cite{voigt1910lehrbuch} is a $6 \times 6$ matrix symmetric about the main diagonal, reflecting the fact that the elasticity tensor has 21 independent components.

The intrinsic material symmetry of a crystal can further reduce the number of independent components \cite{nye1985physical,mouhat2014necessary}.
For example, copper is a cubic crystal with mirror planes and three-fold rotation axes (point group m$\bar3$m), and such symmetry results in a number of only three independent components.
Formally, the material symmetry imposes the following constraints on the elasticity tensor \cite{forte1996symmetry,tadmor2012continuum}:
\begin{equation}\label{eq:mat:sym}
    C_{ijkl} = Q_{ip} Q_{jq} Q_{kr} Q_{ls} C_{pqrs},
\end{equation}
where $\bm Q \in G \subset SO(3)$, and $G$ is the material symmetry group of the crystal, which is a subgroup of the rotation group $SO(3)$.
An interesting question is: how many unique symmetry classes exist and what is the number of independent components in each class?

It turns out that there exists only eight distinct classes (\fref{fig:tensor:components}), proved via a purely algebraic approach by directly identifying the equivalence classes corresponding to \eref{eq:mat:sym} \cite{forte1996symmetry}.
Of the eight classes, one is for isotropic materials, and each of the other seven corresponds to a crystal system \cite{chadwick2001new,tadmor2012continuum}.
In our opinion, there is still significant confusion on this topic.
For example, the categorization by Wallace \cite{wallace1972thermodynamics} and populated by Nye~\cite{nye1985physical}, which incorrectly gives two unique classes for each of the tetragonal and trigonal cases (Fig.~S1 in the \gls{si}), is still widely cited in recent works \cite{singh2021mechelastic,li2022elast,ran2023velas}.
We refer to Section 6.5 of \onlinecite{tadmor2012continuum} for a historical note on the development of the categorization.

The Voigt matrix provides a visual way to observe the material symmetry of a crystal reflected in the elasticity tensor (\fref{fig:tensor:components}).
The values of the matrix components, however, depend on the choice of the coordinate system and do not show any systematic pattern upon coordinate transformation \cite{itin2013constitutive,itin2020irreducible}, making it difficult to build predictive models for elasticity tensors.
This can be overcome by the \emph{harmonic decomposition} \cite{backus1970geometrical}, where the space of all elasticity tensors is factored into the direct sum of irreducible representations of $SO(3)$.
Consequently, any elasticity tensor can be written in the form,
\begin{equation}\label{eq:C:decomp}
    \bm C = h_1 (\lambda)  + h_2(\eta) + h_3 (\bm A) + h_4 (\bm B) +  h_5(\bm H) ,
\end{equation}
where $\lambda$ and $\eta$ (scalars) are the \emph{isotropic} part, $\bm A$ and $\bm B$ (second-rank traceless symmetric tensors) are the \emph{deviatoric} part, and $\bm H$ (fourth-rank traceless symmetric tensor) is the \emph{harmonic} part \cite{backus1970geometrical}.
The harmonic decomposition has two advantageous characteristics.
First, for a given $\bm C$, the values of $\lambda, \eta,\bm A, \bm B, \bm H $ and the functions $h_1, \dots, h_5$ are uniquely determined \cite{backus1970geometrical,forte1996symmetry} (see \gls{si} for their expressions).
Second, each part in \eref{eq:C:decomp} transforms in a known manner with respect to $SO(3)$ operations, enabling the construction of predictive models that leverage recent advancements in geometric deep learning.

\subsection{Equivariant graph neural networks for high-rank tensors}

\begin{figure}[t!]
    \centering
    \includegraphics[width=0.8\columnwidth]{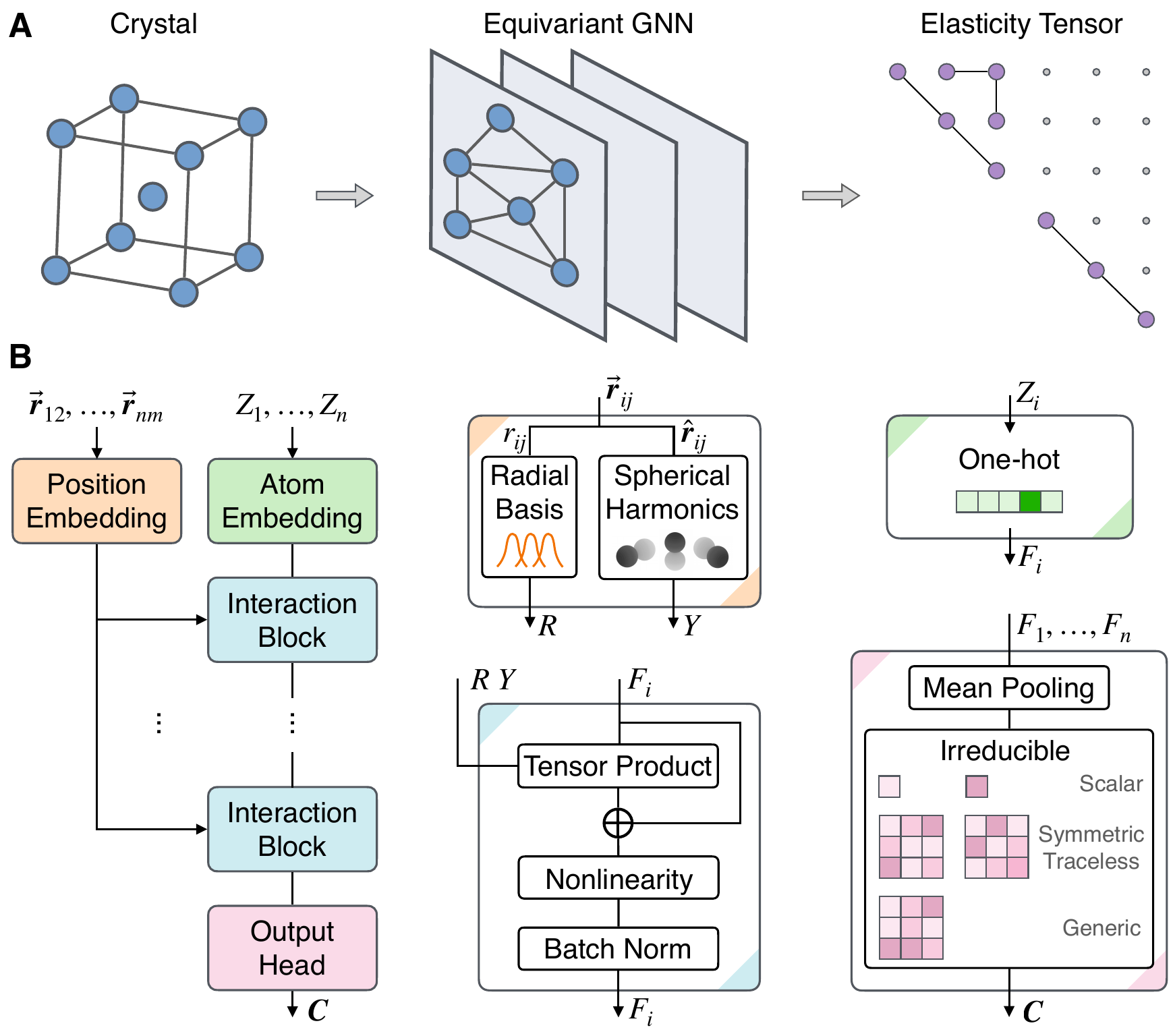}
    \caption{Schematic overview of the \net graph neural network model. (A) The model takes a crystal structure as input, performs message passing with equivariant graph neural network (GNN) layers, and outputs the full elasticity tensor satisfying all symmetry requirements.
        (B) Architecture of the model.
        The model consists of four modules:
        the position embedding module converts an input displacement vector $\vec{\bm r}_{ij}$ between atoms $i$ and $j$ to radial ($R$) and spherical ($Y$) expansions;
        the atom embedding module encodes the atomic number $Z_i$ as irreducible atom features $F_i$ using a one-hot encoding;
        interaction blocks iteratively refine the features using convolutions based on tensor product;
        and the output head selects the appropriate irreducible features corresponding to the elasticity tensor and assembles them as the output tensor $\bm C$.
        Following \onlinecite{itin2020irreducible}, the fourth-rank harmonic part of the elasticity tensor is depicted as a generic matrix.
        The $\oplus$ symbol denotes addition.
    }
    \label{fig:schematic}
\end{figure}

The \net model captures the structure--property relationship of crystalline materials.
It takes a crystal structure as input, represents it as a three-dimensional crystal graph, performs feature updates on the crystal graph, and finally outputs a tensor property of the material built according to the reverse process of the harmonic decomposition in \eref{eq:C:decomp}.

In the GNN model (\fref{fig:schematic}), the input crystal is represented as a graph  $\mathcal{G} (V, E)$,
with atoms as the nodes $V$ and bonds as the edges $E$.
The feature $F_i \in V$ characterizes atom $i$, and the initial value of $F_i$ is obtained by encoding the atomic number $Z_i$ using a one-hot scheme.
A bond/edge between two atoms is created if the distance $\| \vec{\bm r}_{ij} \| $ is smaller than a cutoff value, where $\vec{\bm r}_{ij}$ denotes the distance vector between atoms $i$ and $j$.
Periodic boundary conditions are considered when constructing the bonds, using super cell vectors.
The distance vector $\vec{\bm r}_{ij}$ is separated into two parts:
the unit vector  $\hat{\bm r}_{ij}$ from atom $i$ to atom $j$ and the scalar distance $r_{ij}$ between them.
The former is expanded on real spherical harmonics $Y_m^{l}$, and the latter is
expanded on the Bessel radial basis functions \cite{gasteiger2020dimenet}.
In sum, these embedding modules extract structural information  (coordinates of atoms, atomic numbers, and super cell vectors)  from the crystal and provide them to the interaction blocks.

The interaction blocks iteratively refine the atom features via convolution operations.
The architecture of the interaction block follows the design of Tensor Field Network \cite{thomas2018tensor} and NequIP \cite{batzner2022e3}.
Unlike many existing GNNs for molecules and crystals \cite{xie2018crystal,chen2019graph,wen2020bondnet,wen2022rxnrep} that utilize scalar features, here, the atom feature $F_i$ is a set of scalars, vectors, and high-rank tensors.
Formally, it is a geometric object consisting of a direct sum of irreducible representations of the $SO(3)$ rotation group \cite{thomas2018tensor,batzner2022e3}.
There are two major benefits of using such geometric features.
First, they are incorporated as inductive bias which can improve model accuracy and reduce the amount of training data.
Second, from them, one can easily construct other physical tensors such as the elasticity tensor in this work.

The convolution on these geometric objects is an equivariant function, meaning that if the input atom feature $F$ to the convolution is transformed under a rotation in $SO(3)$, the output is transformed accordingly.
This is achieved via the tensor product convolution by updating the atom feature in the $(k+1)_\mathrm{th}$ interaction block from that in the $k_\mathrm{th}$ interaction block,
\begin{equation}\label{eq:convolution}
    F_i^{k+1} =  \sum_{j\in \mathcal{N}_i} R(r_{ij}) Y(\hat{\bm r}_{ij})\otimes F_j^{k},
\end{equation}
where $\mathcal {N}_i$ denotes the set of neighboring atoms for atom $i$, $R$ indicates a multilayer perceptron (MLP) on the radial basis expansion of $r_{ij}$, and $Y$ indicates the spherical harmonics expansion of $\hat{\bm r}_{ij}$.
The tensor product $\otimes$ between $Y$ and $F_j^k$ is a bilinear map, which is a generalization of the outer product of two vectors.
The product output is decomposed back onto the irreducible representations, and the entire operation is equivariant
\cite{thomas2018tensor}.
The use of an MLP makes the convolution learnable.
After a skip connection \cite{he2016deep}, the feature $F_i^{k+1}$ is passed through a nonlinear activation function and finally normalized using an equivariant normalization function \cite{e3nnpaper}.

The output head maps the refined features from the interaction blocks to the target materials tensor of interest.
First, the features of all atoms are aggregated to obtain a representation of the crystal.
For intensive properties such as the elasticity tensor, meaning that the property value does not depend on the size of the system, we adopt the mean pooling by averaging the features such that the representation of the crystal is independent of the number of atoms.
Next, an appropriate subset of the pooled irreducible representations that correspond to the target tensor of interest is selected and then assembled as the model output.
For the elasticity tensor, the selected ones are two scalars, two second-rank traceless symmetric tensors, and a fourth-rank harmonic tensor.
They are assembled to an elasticity tensor according to \eref{eq:C:decomp}.

Overall, \net is a function $\bm C = f(x)$ that maps a crystal structure $x$ to its elasticity tensor $\bm C$.
The function $f$ is equivariant to the $SO(3)$ group transformation, that is,
for any $x$ and $g \in  SO(3)$, we have $ D_y(g) f(x) = f( D_x(g) x) $, where $D_x(g)$ and $D_y(g)$ are rotation matrices parameterized by $g$ for the crystal structure and the elasticity tensor, respectively.
This ensures that the model can produce an elasticity tensor $\bm C$ that respects the orientation of the input crystal structure.
In other words, the choice of a specific coordinate system does not affect the model output; if the coordinate system is rotated, the output tensor rotates accordingly.
This \emph{independence of the frame of reference} characteristic is an indispensable property for models that predict tensors.
In addition, any such model should also \emph{preserve the material symmetry} of the crystal.
By construction, \net guarantees the material symmetry reflected in the elasticity tensor.
Concretely, if the predicted elasticity tensor is represented as a Voigt matrix, the symmetry and number of independent components in \fref{fig:tensor:components} are automatically maintained for all seven crystal systems (proof in the \gls{si}).
For example, for a cubic crystal, the model guarantees that there are only three independent components $c_{11}=c_{22}=c_{33}$, $c_{12}=c_{13}=c_{23}$, and $c_{44}=c_{55} =c_{66}$ and that all other components are zero.

\subsection{Elastic properties of polycrystals}
\label{sec:scalar:prop}

\begin{figure}[b!]
    \centering
    \includegraphics[width=1\columnwidth]{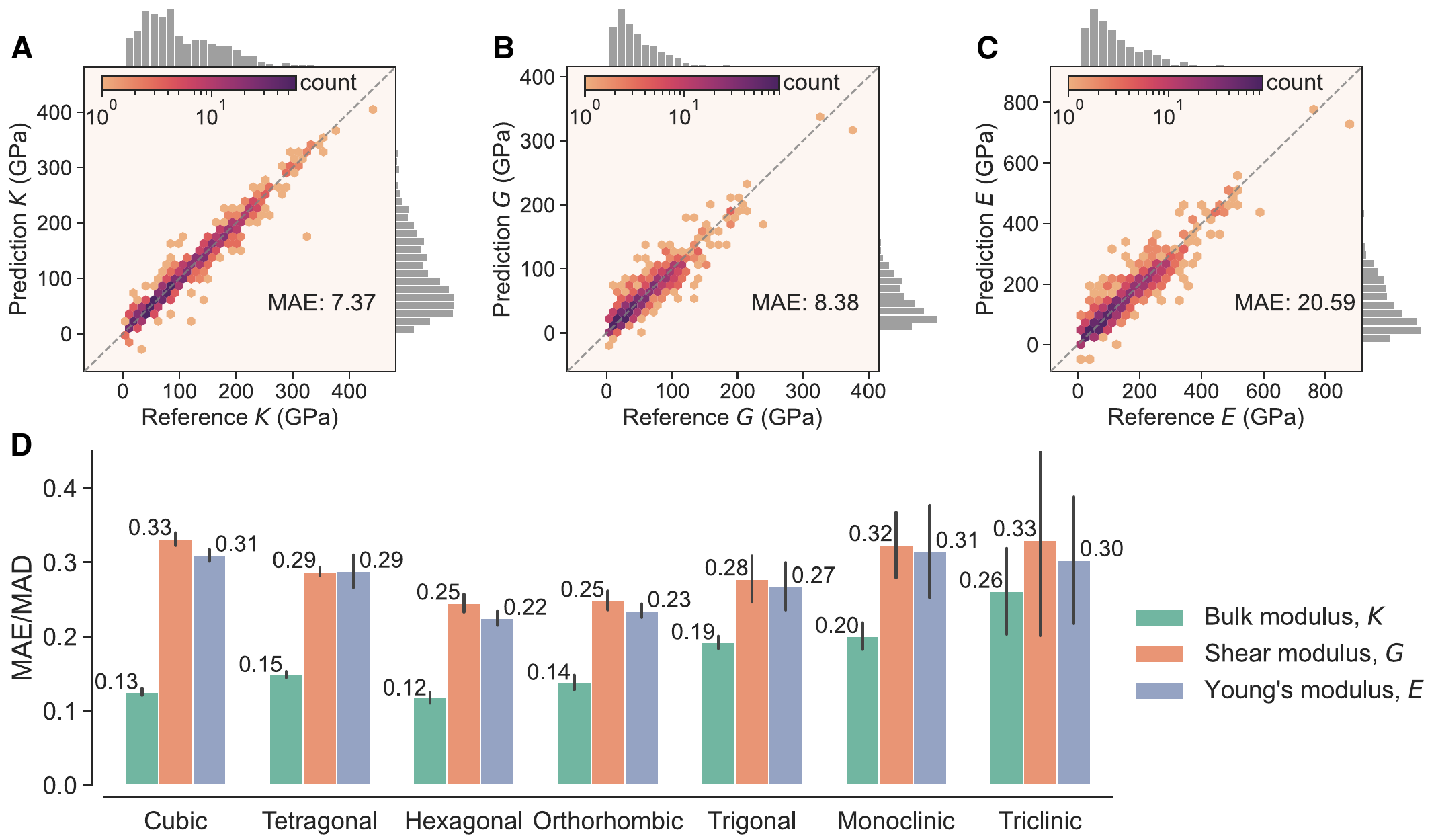}
    \caption{Performance of \net on various elastic properties.
        (A) Bulk modulus, (B) shear modulus, and (C) Young's modulus predicted by \net versus DFT reference values.
        (D) Scaled error by crystal systems. The error bar indicates the standard deviation obtained from an ensemble of five models trained with different initializations.
        MAE: mean absolute error; MAD: mean absolute deviation.
    }
    \label{fig:error:scalar}
\end{figure}

The \net model directly outputs the full elasticity tensor.
To assess its performance, we computed several commonly used elastic properties for polycrystals from the elasticity tensor.
\fref{fig:error:scalar} illustrates the results on the moduli obtained using the Hill average scheme \cite{hill1952elastic}.
The DFT reference values have a range of  $4\sim442$~GPa,  $3\sim375$~GPa, and $9\sim878$~GPa for the bulk modulus $K$, shear modulus $G$, and Young's modulus $E$, respectively.
The formulas to obtain the moduli are given in \sref{sec:methods}, and their statistics and distribution are given in Figs.~S4--S7 in the \gls{si}.
The predictions of \net closely align with the DFT reference values along the entire ranges, achieving mean absolute errors (MAEs) of 7.37~GPa, 8.38~GPa, and 20.59~GPa for $K$, $G$, and $E$, respectively.
To connect the MAE values to practical applications, let's examine the error in strain caused by the MAE in Young's modulus.
For example, at $E=128.4$~GPa (the mean of DFT references), an error of $20.59$~GPa will lead to a relative error of 19\% in the strain (calculation given in the \gls{si}).
While different applications necessitate varied accuracy, a relative error of 19\% in the strain can be acceptable, given that noncontact experimental techniques for strain measurement such as the digital image correlation method \cite{reu2018dic} have a typical error of $\sim$10\%.

For comparison, we trained two additional models.
The first is a variant of \net, where the tensor output head of \net is replaced with a scalar output head, referred to as MatSca hereafter.
The second is the AutoMatminer algorithm, an automated machine learning pipeline designed for predicting scalar materials properties \cite{dunn2020benchmarking}.
We evaluated their performance in predicting the elastic moduli, and the results are listed in \tref{tab:mae:mad}.
Both \net and MatSca have smaller MAEs than AutoMatminer across all three moduli, owning to the effectiveness of the underlying neural networks in learning materials properties from structures.
The performance of \net and MatSca are comparable.
However, it is worth highlighting that while an individual MatSca model was trained for each modulus, a single \net model successfully produced all the elastic moduli, demonstrating the versatility of the \net model.

Upon closer examination of \fref{fig:error:scalar}A--C, we have identified some inconsistencies in the predictions.
All crystals in the DFT dataset have positive moduli, the predicted moduli by \net, however, occasionally yield negative values, indicating that the associated crystal is elastically unstable.
This is an inherent challenge faced by machine learning regression models in general, albeit with physical inductive biases embedded in the model such as the symmetry requirements in \net.
The number of crystals with negative predicted moduli remains minimal, accounting for only 3, 2, and 2 out of the 1021 test data for bulk, shear, and Young's moduli, respectively.
The moduli alone, however, do not provide a comprehensive understanding.
For a crystal to be elastically stable, the sufficient and necessary condition is that the Voigt matrix should be positive definite \cite{mouhat2014necessary}.
We checked this and found that 25 crystals in the test set do not satisfy this condition.
The majority of them are due to the incorrect prediction of the relative magnitudes of the diagonal and off-diagonal components of the Voigt matrix.
A breakdown of the errors is provided in the \gls{si}.
Nevertheless, this is not a concern in practical use; one can filter out the negative ones if desired.

\begin{table}
    \caption{Performance of the models in predicting the bulk, shear, and Young's
        moduli.
        The value in a pair of parentheses is the standard deviation obtained from an ensemble of five models trained with different initializations.
        MAE: mean absolute error; MAD: mean absolute deviation.
    }
    \label{tab:mae:mad}
    \centering
    \begin{tabular}{@{\extracolsep{5pt}}ccccccc}
        \hline
                     & \multicolumn{2}{c}{$K$ (GPa)}
                     & \multicolumn{2}{c}{$G$ (GPa)}
                     & \multicolumn{2}{c}{$E$ (GPa)}                                                                              \\
        \cline{2-3} \cline{4-5} \cline{6-7}
                     & MAE                           & MAE/MAD       & MAE         & MAE/MAD       & MAE          & MAE/MAD       \\
        \hline
        \net         & 7.37 (0.10)                   & 0.130 (0.002) & 8.38 (0.16) & 0.280 (0.005) & 20.59 (0.35) & 0.275 (0.005) \\
        MatSca       & 7.32 (0.09)                   & 0.129 (0.002) & 8.63 (0.07) & 0.288 (0.002) & 19.87 (0.43) & 0.265 (0.006) \\
        AutoMatminer & 9.84 (0.34)                   & 0.174 (0.006) & 9.27 (0.32) & 0.309 (0.011) & 22.10 (0.77) & 0.295 (0.024) \\
        \hline\end{tabular}
\end{table}

To assess how \net performs for different elastic properties as well as for different crystal systems, we computed the scaled error, $\text{SE} = \text{MAE}/\text{MAD}$, in which MAE and MAD are the mean absolute error and mean absolute deviation, respectively (see \sref{sec:methods}).
MAD quantifies the distance of reference values to their mean, and larger MAD means the reference values are more scattered.
A model that makes accurate predictions for each data point will have an SE of 0, and a model that always predicts the mean of the dataset will have an SE of exactly 1.
Comparing between properties, we see from \fref{fig:error:scalar}d that the SE of bulk modulus is smaller than those of shear modulus and Young's modulus across all crystal systems.
This suggests that bulk modulus is easier to predict, in agreement with existing observations \cite{chen2019graph,dunn2020benchmarking,wang2021compositionally,de2021materials}.
Next, we compare between crystal systems.
The dataset used for model development has an uneven distribution in terms of the number of materials in each crystal system (Fig.~S2 in the \gls{si}).
For example, it contains 4217 cubic crystals, fewer than 800 trigonal and monoclinic crystals, and only 60 triclinic crystals.
As a result, the SE and the error bar of the predicted moduli are larger for trigonal, monoclinic, and triclinic crystals in general (\fref{fig:error:scalar}d).
Despite the slightly higher errors, it is notable that the model can still perform well for the crystal systems with a low presence in the training data, particularly for triclinic crystals.
This is primarily because \net internally treats all crystals the same, enabling crystal systems with fewer data to leverage the abundant data from other crystal systems and acquire enhanced representations.
This type of transferability is not possible with models that are built separately for each crystal system.

The elastic moduli have values across different orders from near zero to hundreds (Figs.~S4--S7 in the \gls{si}).
To mitigate the challenge of learning values across a broad spectrum, some existing models  \cite{dunn2020benchmarking,chen2019graph,wang2021compositionally,de2021materials}
adopt the approach of predicting the logarithm of the moduli.
Unlike these models, \net directly predicts the full tensor without any data transformation, and all other elastic properties (including their logarithms) can be computed from it.
The logarithms of the bulk, shear, and Young's moduli obtained from \net are comparable to those that learn in the logarithmic space
(Table~S1 in the \gls{si}).
Further, the performance of \net can be further improved with additional training data.
The MAE almost decreases linearly with the logarithm of the number of data used to train the model (Fig.~S10 in the \gls{si}).
Finally, it is also possible to predict the full elasticity tensor by separately modeling its non-zero independent components \cite{ahmad2018machine, revi2021machine}.
\net performs much better than this approach thanks to its ability to
deal with all crystal systems within a united framework (further discussion in the \gls{si}).

\subsection{Anisotropic elastic properties}
\label{sec:3d:prop}

\begin{figure}[b!]
    \centering
    \includegraphics[width=1\columnwidth]{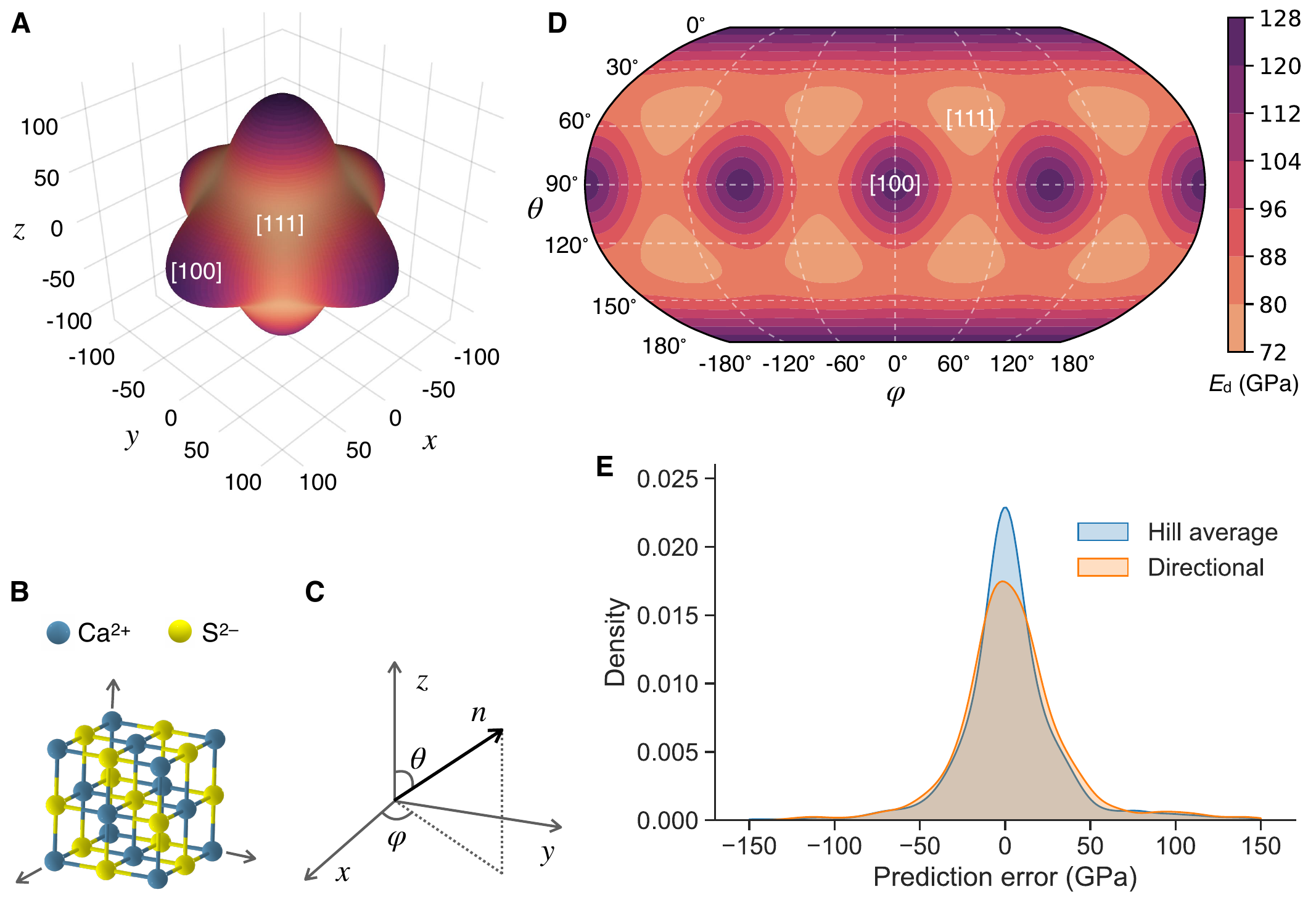}
    \caption{Model performance on the directional Young's modulus $E_\text{d}$.
        (A) Three-dimensional representation of Young's modulus predicted by \net,
        for (B) the CaS rocksalt crystal.
        With (C) a spherical coordinate system and the Robinson map projection,
        (D) $E_\text{d}$ is represented in two dimensions.
        (E) Distribution of the prediction error in $E_\text{d}$; also included is the prediction error in the isotropic Young's modulus from Hill average.
    }
    \label{fig:3d:E}
\end{figure}

Crystals are inherently anisotropic, and thus their elastic properties can vary depending on the direction of measurement.
This anisotropy arises from a crystal's structure, including the symmetry of the lattice and the arrangement of the atoms.
\net predicts the full elasticity tensor, and, for the first time with a machine learning model, we are able to investigate the anisotropic elastic behavior of crystals.
We focus our discussion on the directional dependence of Young's modulus (further results on shear modulus given in the \gls{si}).

Young's modulus $E$ discussed in \sref{sec:scalar:prop} is an averaged property for isotropic polycrystals.
But for single crystals, Young's modulus depends on the direction along which the strain/stress is applied and measured.
Given the elasticity tensor $C_{ijkl}$ (equivalently, the compliance tensor $S_{ijkl}$, see \sref{sec:methods}), the directional Young's modulus can be computed as \cite{nye1985physical,ran2023velas}
\begin{equation}\label{eq:Ed}
    E_\text{d} ({\bm n})  =(n_i n_j n_k n_l S_{ijlk})^{-1},
\end{equation}
where $\bm n$ is a unit vector that specifies the direction of measurement.
The direction dependence of Young's modulus can be visualized with a three-dimensional plot (\fref{fig:3d:E}a).
Interactive visualization can be obtained via, for example, the \verb|elate| package \cite{gaillac2016elate}.
Alternatively, via a spherical coordinates transformation:
$\bm n = [\sin\theta\cdot\cos\varphi, \sin\theta\cdot\sin\varphi, \cos\theta]$ (\fref{fig:3d:E}c), it can be represented in two dimensions (\fref{fig:3d:E}d, with a Robinson map projection \cite{robinson1974new}).
Such plots make it easier to visually investigate the anisotropic characteristics of Young's modulus.
For example, for the cubic rocksalt CaS crystal (\fref{fig:3d:E}b), the maximum directional Young's modulus $E_\text{d}^\text{max}$ is along the $\langle 100 \rangle$ directions (for instance, $\theta = 90^\circ$ and $\varphi = 0^\circ$),
while the minimum $E_\text{d}^\text{min}$ is along the the $\langle 111 \rangle$ directions (for instance, $\theta = 54.7^\circ$ and $\varphi = 45^\circ$).
In fact, for cubic crystals such as CaS, the extreme values of $E_\text{d}$ are guaranteed to occur in these two directions \cite{nye1985physical}.
\eref{eq:Ed} can be simplified as $E_\text{d}(\bm n) = [S_{1111} - 2(S_{1111} - S_{1122} - 2S_{2323})(n_1^2n_2^2 + n_2^2n_3^2 + n_3^2n_1^2)]^{-1}$ for cubic crystals, expressed in terms of their three independent elasticity tensor components.
It can be mathematically shown that, if
\begin{equation}\label{eq:E:max:dir}
    S_{1111} - S_{1122} - 2S_{2323} < 0,
\end{equation}
Young's modulus achieves its maximum $E_\text{d}^\text{max}$ in the $\langle 100 \rangle$ directions and minimum in the $\langle 111 \rangle$ directions;
otherwise, if $S_{1111} - S_{1122} - 2S_{2323} > 0$, the two extremes switch directions (derivation given in the \gls{si}).
\eref{eq:E:max:dir} is satisfied by CaS, and thus we observe the maximum and minimum in the $\langle 100 \rangle$ and $\langle 100 \rangle$ directions, respectively.
In addition, $E_\text{d}$ of CaS possesses symmetry (for example, 3-fold rotational axis along the cube diagonals) consistent with a cubic crystal, further confirming that the predicted elasticity tensor \emph{preserves material symmetry}.

To quantitatively assess the ability of \net in predicting anisotropic elastic properties, we measured the error between the model predicted directional Young's modulus $E_\text{d}^\text{pred}$ and the DFT reference $E_\text{d}^\text{ref}$.
For CaS, \net prediction closely follows DFT reference,
with a maximum under-prediction of $8.7$~GPa along the $\langle 111 \rangle$ directions and a maximum over-prediction of 9.4~GPa along the $\langle 100 \rangle$ directions (Fig.~S14 in the \gls{si}).
In addition to this example crystal, we calculated the error for the entire test set, computed as $D = D_s \cdot D_v$ for each crystal.
The value of the error is $ D_v = \int_\theta\int_\varphi  \vert \Delta \text{E}_d \vert \, \text{d}\theta \text{d}\varphi$, where $ \Delta E_d (\theta, \varphi) = E_\text{d}^\text{pred} (\theta, \varphi)  - E_\text{d}^\text{ref} (\theta, \varphi)$ and $\vert\cdot\vert$ denotes the absolute value.
The sign of the error is $D_s=+1$ if$  \int_\theta\int_\varphi  \Delta \text{E}_d  \, \text{d}\theta \text{d}\varphi > 0 $, and $D_s = -1$ otherwise.
Put differently, the value $D_v$ quantifies the average deviation from the DFT reference, while the sign $D_s$ characterizes whether the overall prediction is larger than the DFT reference. The integration over $\theta$ and $\varphi$ is performed using the Chebyshev quadratures, which uniformly distribute the integration points on the sphere and can avoid biasing specific points \cite{beentjes2015quadrature}.
The distribution of the prediction error $D$ of $E_\text{d}$ for the test set is plotted in \fref{fig:3d:E}e.
It has a Gaussian-like shape, and it almost overlaps with that of isotropic Young's modulus obtained using the Hill average.
Similar behaviors are observed for shear modulus and MAEs by crystal systems (Figs.~S15 and S16 in the \gls{si}).
These observations suggest that, on average, \net performs equally well for the anisotropic and averaged isotropic elastic properties.
Accurate prediction of anisotropic properties offers a comprehensive understanding of the elastic characteristics exhibited by crystals, enabling the discovery of new materials through the utilization of these predictive capabilities.

\subsection{Screening of crystals with extreme properties}

\begin{figure}[htb!]
    \centering
    \includegraphics[width=0.55\columnwidth]{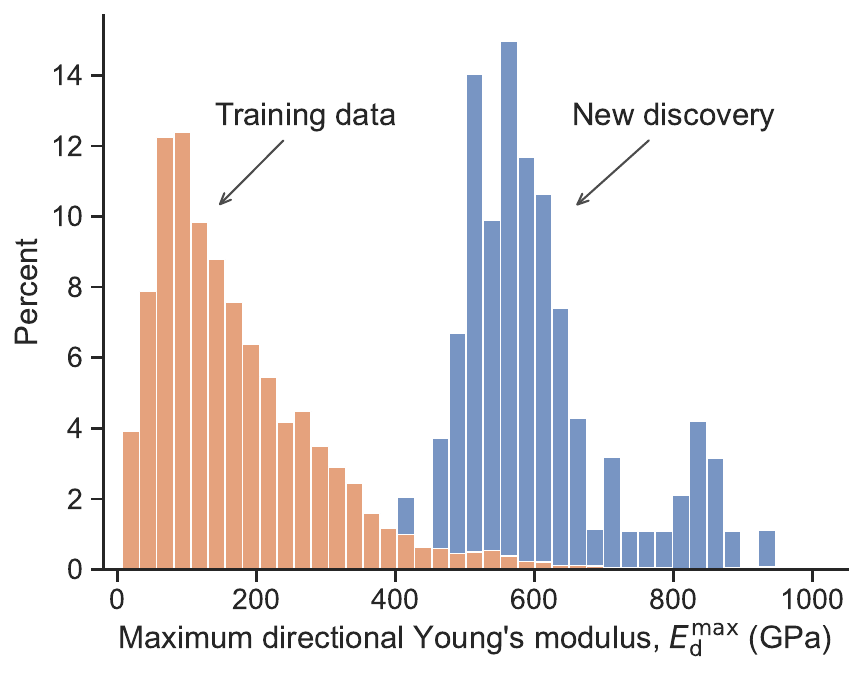}
    \caption{Screening of crystals with large maximum directional Young's modulus. The training data and new discoveries are separately normalized such that each sums to 100 percent.
    }
    \label{fig:max:E}
\end{figure}

The maximum of the directional Young's modulus, $E_\text{d}^\text{max}$, characterizes the smallest possible deformation of a crystal under applied external loading.
It helps in the selection and orientating of materials to minimize shape change to guarantee the reliability of precision devices such as micro-electromechanical systems \cite{huang2012mems}.

We have also applied the \net model to screen for crystals with large  $E_\text{d}^\text{max}$.
We first filtered crystals from the Materials Project database based on their energy above the convex hull values, selecting those with a value of $\leq 50$~meV/atom.
This energy determines the thermodynamic stability of a crystal and has been shown to correlate with the synthesizability of crystals \cite{sun2016thermodynamic,bartel2022review}.
We further narrowed down the selection to crystals with fewer than 50 atomic sites to reduce computational cost and remove crystals already present in the dataset used to develop the model.
This resulted in 53480 crystals for further analysis.
Next, we employed \net to predict the full elasticity tensors for these crystals and compute their $E_\text{d}^\text{max}$.
The top 100 crystals with the highest  $E_\text{d}^\text{max}$ were then supplied for DFT computation to obtain their elasticity tensors.
\fref{fig:max:E} presents a histogram of $E_\text{d}^\text{max}$ for the identified 100 crystals.
Their $E_\text{d}^\text{max}$ values all fall at the tail of the distribution of $E_\text{d}^\text{max}$ from the training data.
Quantitatively, the mean of $E_\text{d}^\text{max}$ for the identified crystals is 606~GPa, while that for the training data is 174~GPa, corresponding to the expected value for a randomly selected material.
This demonstrates the effectiveness of leveraging \net to screen for materials with extreme elastic properties.
The newly identified crystals are provided in Data Availability.

In addition, we have identified 11 unconventional polymorphs of elemental cubic metals regarding the direction of the extreme values of Young's modulus.
Five of them have already been experimentally synthesized (four stable ground-state polymorphs and one metastable polymorph \cite{bartel2022review}), and the other six are metastable polymorphs that have not yet been experimentally observed.
It is believed that $E_\text{d}^\text{max}$ is along the $\langle 111 \rangle$ directions (meaning \eref{eq:E:max:dir} is not satisfied) for all cubic metals except Mo \cite{nye1985physical}.
We suspect that there exist other polymorphs of elemental cubic metals satisfying the criterion in \eref{eq:E:max:dir}.
To test this, we performed a screening of the Materials Project database.
From the dataset used to develop \net, we have identified six such crystals.
Mo is among them, and the other five are V (one polymorph), Cr (two polymorphs), and W (two polymorphs).
For crystals not in the dataset, we first used \net to predict their elasticity tensors and then selected the 18 crystals that meet the verification criterion using DFT.
DFT predicted 12 crystals satisfying the criterion; six are elastically unstable structures whose Voigt matrix has negative eigenvalues \cite{mouhat2014necessary, tolborg2022free},
and the other six successful ones are Mn, Na, K, Cs, Rh, and Tl (one polymorph for each).
Among the 11 newly identified crystals, polymorphs of alkali metals Na, K, and Cs have DFT calculated $S_{1111} - S_{1122} - 2S_{2323}$ values much more negative than that of Mo,
but they are all hypothetical crystals that have not been experimentally synthesized yet.
Five polymorphs are indeed experimentally observed, and they are all neighbors of Mo in the periodic table, namely V, Cr, Mn, and W.
Among them, four are thermodynamically stable ground-state polymorphs, and one polymorph of Cr is metastable.
Crystal structures, ground-state information, and elasticity tensors of the 11 unconventional polymorphs are provided in Data Availability and Table~S3 in the \gls{si}.

\section{Conclusions}
\label{sec:discussion}

A model such as \net that can predict the full elasticity tensors of inorganic crystalline compounds across crystal systems and chemical species brings new possibilities to probe and design materials with targeted mechanical properties.
\net has several unique characteristics:
1).\ it learns the full elasticity tensor and automatically handles all symmetry requirements, without the need to build separate models for individual components of the tensor or for each crystal system;
2).\ any elastic properties such as the bulk, shear, and Young's moduli can be computed from the predicted elasticity tensor, leading to a unified framework for modeling elasticity; and
3).\ it allows for the exploration of anisotropic elastic behaviors (not possible with existing machine learning models), demonstrated by screening for crystals with extreme directional Young's modulus.

It should be noted, however, that a crucial aspect regarding the practical use of the model relies on the robustness of the input structure.
Given two structures of a crystal where the atomic coordinates are slightly different, we would expect the elastic properties to be similar.
If, otherwise, the model is extremely sensitive to the input, then it is not ideal for practical application.
Extra work such as highly accurate DFT structure relaxation is needed before applying the model for predictions.
We tested \net by using structures directly queried from the Materials Project database and structures with tighter geometry optimization.
The MAE of $E_\text{d}^\text{max}$ between using the two types of structures is 6.55~GPa (Fig.~S17 in the \gls{si}), more than three times smaller than the MAE (22.36 GPa) of $E_\text{d}^\text{max}$ between model predictions and DFT references.
This suggests that \net is robust enough to its input, and reasonably optimized structures (for example, from online databases) would not introduce extra error larger than the intrinsic error in the model.

The \net model is not limited to inorganic crystalline compounds and even elasticity tensors in general.
The elastic behavior of other classes of materials such as two-dimensional layered materials and molecular crystals play a significant role in determining their functionality, and \net can be directly applied to model their elasticity tensors.
Of course, a curated dataset of reference elasticity tensors is needed.
Such data already exists, for example, in the Computational 2D Materials Database (C2DB) \cite{gjerding2021recent}.
Moreover, besides elasticity tensor, \net can be applied to other tensorial properties of materials.
These can be broadly categorized into two classes: material-level property and atom-level property.
While the former means a single tensor for each crystal, the latter means a separate tensor for each atom in the crystal.
Other material-level properties such as piezoelectric
and dielectric tensors can be modeled by updating the output head as in \fref{fig:schematic} to use the corresponding irreducible representations of the tensor of interest (for example, a single second-rank symmetric matrix for the dielectric tensor).
For atom-level properties such as the neutron magnet resonance (NMR) tensor, instead of using a mean pooling to aggregate atom features, one can directly map the features of an atom to a tensor for that atom.
Using \net, we have conducted such an analysis for NMR tensors of silicon oxides and found that \net significantly outperforms both historic analytical models and other machine learning models
by more than 50\% for isotropic and anisotropic NMR chemical shift \cite{venetos2023machine}.

One potential limitation of the proposed approach is the reliance on a relatively large dataset to develop the model.
We have curated a dataset of 10276 elasticity tensors which took millions of CPU hours to obtain.
Such large datasets for other tensorial properties may not be readily available, but they begin to emerge.
For example, the Materials Project has about 3000 piezoelectric and 7000 dielectric tensors \cite{de2015database,petousis2017high}.
This amount of data might still be a good start to training faithful models, given that piezoelectric and dielectric are third- and second-rank tensors, respectively, which are much simpler than the fourth-rank elasticity tensor.
In fact, for the second-rank NMR tensor, we only used a dataset of 421 crystals to obtain the best-performing model \cite{venetos2023machine}.
Another possibility is to apply a transfer learning approach, where the model is first trained on a different property with large data (for instance, elasticity tensor) and then finetuned on the target property of interest (for instance, piezoelectric tensor).
A limitation of the trained model can come from the data.
The data consists of DFT calculations of perfect single crystals with relatively small super cells at a temperature of 0~K.
Given that the efficacy of the model is intrinsically tied to the scope of the training data,
it is imperative to exercise caution when applying the model
to scenarios that extend beyond these parameters.
For example, the model is not appropriate for crystals with defects, such as vacancies, dislocations, and grain boundaries.
Additionally, it is not advisable to directly employ the model for estimating the mechanical properties at finite temperate, especially for those materials, like metallic alloys, which exhibit a pronounced temperature dependency.

\section{Experimental}
\label{sec:methods}

\subsection*{Data generation}

The elasticity tensors were computed by a liner fitting of the stresses and strains obtained from DFT calculations using the Vienna Ab Initio Simulation Package (VASP) \cite{kresse1993ab}.
The calculations follow the same procedures discussed in \onlinecite{de2015charting}, using \verb|PREC=Accurate|, a tight convergence criterion of \verb|EDIFF=1e-6|, an energy cutoff of \verb|ENCUT=700 eV|, and a $k$-points density of 64~\AA$^{-3}$ in the reciprocal space to sample the Brillouin zone.
Two additional improvements are made.
First, to get more precise stresses for calculating the elasticity tensor, the projection operators in VASP are evaluated in the reciprocal space, that is, the setting \verb|LREAL=False| was adopted.
Second, to reduce numerical error in the calculations, the stresses are symmetrized according to the crystal symmetry.
The entire workflow was implemented in the open-source \verb|atomate| package \cite{mathew2017atomate}.

\subsection*{Model architecture}

A crystal structure is converted to a graph using a distance-based approach, where an edge is created between a pair of atoms if their distance is smaller than a cutoff radius $r_\text{cut}$.
Periodic boundary conditions are considered in the graph construction.

For atom $a$, its atomic number $Z_a$ is embedded as a vector with $c$ components
using a one-hot encoding to obtain the initial atom feature $F_a$.
The unit vector $\hat{\bm r}_{ij}$ from atom $i$ to atom $j$ is expanded using a spherical harmonics basis consisting up to a degree of $l=4$.
(Explicitly, this corresponds to the ``0e + 1o + 2e + 3o + 4e'' irreducible representations in \verb|e3nn| notation \cite{e3nnpaper}).
The distance $r_{ij}$ between atoms $i$ and $j$ is expanded into a vector $R$ using the radial basis functions \cite{gasteiger2020dimenet},
\begin{equation}
    \text{RBF}_n(d) =  \sqrt{\frac{2}{r_\text{cut}}} \frac{\sin(\frac{n\pi}{r_\text{cut}} d)}{d}
\end{equation}
where $n = 1,2,\dots,$ is an index of the radial basis.

With the atom features $F$, the spherical harmonics expansion of $\hat{\bm r}_{ij}$, and the radial basis expansion $R$ of $r_{ij}$ obtained from the embedding layers, the interaction block performs tensor product--based convolution to refine the atom features.
This is achieved via \eref{eq:convolution}, more specifically \cite{thomas2018tensor},
\begin{equation} \label{eq:conv:indicial}
    \mathcal{L}_{acm_o}^{l_o} (\{\vec{\bm r}_{ab}\}, \{F^{l_i}_{bcm_i}\})
    = \sum_{m_i, m_f}C^{(l_o,m_o)}_{(l_f, m_f)(l_i, m_i)}
    \sum_{b\in \mathcal{N}_a}
    R_c^{(l_f, l_i)}(r_{ab}) Y_{l_f}^{m_f}(\hat{\bm r}_{ab})F^{l_i}_{bcm_i},
\end{equation}
where $a$ denotes the center atom, $b$ denotes all its neighbors $\mathcal{N}_b$ within the cutoff $r_\text{cut}$;
$l$ is an integer indicating the degree of the spherical harmonic function, and $m = -l, \dots, l$;
the subscripts $i$, $o$, and $f$ indicate input, output, and filter, respectively;
$c$ is the channel index (for example, for the embedding layer, it indicates the components of the one-hot encoding);
$C$ denotes the Clebsch-Gordan coefficients;
and finally, $R_c^{(l_f, l_i)}$ are learnable multilayer perceptrons (MLPs), which take the RBF expansion as the input and contain most of the parameters of the model.
Essentially, this combines the atom features of neighbors $b$ to be the new atom features of the center atom $a$, in the same spirit of a message-passing graph neural network.
A major characteristic of \eref{eq:conv:indicial} is that the use of the spherical harmonics and the Clebsch-Gordan coefficients together imply that convolutions are equivariant \cite{thomas2018tensor}.

The atom features $F$ is also passed through a self-interaction,
\begin{equation}
    \mathcal{S}_{acm}^l(F_{ac'm}^l) =  \sum_{c'} W_{cc'}^l F_{ac'm}^l ,
\end{equation}
where $W_{cc'}^l$ are learnable model parameters.
The updated atom features are then obtained as
\begin{equation}
    F^{k+1} =\mathcal{L}(F^k) + \mathcal{S}(F^k),
\end{equation}
where $F^{k}$ denotes the features in interaction block $k$, and $F^{k+1}$ the features in the next interaction block.
Indeed, $F^{k+1}$ are further passed through a nonlinearity and a normalization functions.
For each scalar part $s$ in $F$, the nonlinearity is chosen to be the SiLU function \cite{hendrycks2016gaussian},
\begin{equation}
    \text{SiLU}(s) = \sigma(s)s,
\end{equation}
and for each non-scalar part $\bm t$ in $F$, the gated nonlinearity \cite{weiler20183d} is adopted,
\begin{equation}
    G(\bm t) =   \sigma(x) \bm t,
\end{equation}
where $\sigma$ is the sigmoid function, and $x$ is a scalar obtained from  \eref{eq:conv:indicial} by setting $l_o = 0$ and $m_o=0$.
Finally, the equivariant batch normalization \cite{e3nnpaper} is applied to the features to avoid gradient vanishing or exploding.

The readout head aggregates the features of individual atoms to obtain a representation of the material via a mean pooling,
\begin{equation}
    F_\text{mat} = \frac{1}{N}\sum_a^N F_a,
\end{equation}
where $F_a$ denotes the features of atom $a$ for a crystal of a total number of $N$ atoms.
From $F_\text{mat}$, the appropriate irreducible representations (``2x0e + 2x2e + 4e'' in \verb|e3nn| notation) that correspond to the elasticity tensor (two scalars, two second-rank traceless symmetric tensors, and one fourth-rank traceless symmetric tensor) are selected to construct the elasticity tensor according to \eref{eq:C:decomp}.

\subsection*{Model training}

The dataset of 10276 elasticity tensors is split into three subsets for training, validation, and testing with a split ratio of 8:1:1.
A random split with stratification is adopted where each of the seven crystal systems is separately treated in the split.
The model parameters are optimized using the training set, model hyperparameters are determined based on model performance on the validation set,
and error analysis is performed using the test set unless otherwise stated.
We train the model with the Adam optimizer to minimize a mean-squared-error loss function
$ L = \sum_i^B \| \bm C_i - \bm C_i^\text{ref} \|^2 $
with a mini-batch size $B$ of 32.
Note, $\bm C_i$ denotes the irreducible representation of the model predicted elasticity tensor with 21 components (see \eref{eq:C:decomp}), but not the Cartesian tensor with 81 components.
Similarly, $\bm C_i^\text{ref}$ denotes the corresponding reference DFT values.
The learning rate is set to 0.01, and a reduce-on-plateau learning rate scheduler is used, which decreases the learning rate by a factor of 0.5 if the validation error does not decrease for 50 epochs.
The training stops when the validation error does not decrease for 150 consecutive epochs, and a maximum of 1000 epochs are allowed for the optimization.
We performed a grid search to obtain model hyperparameters such as the $r_\text{cut}$ and $c$.
Search ranges and their optimal values are listed in Table~S4 in the \gls{si}.

Ten-fold cross validation is also performed to test the effects of different data splits.
Figs.~S11 and S12 in the \gls{si} present the results, and \net is not sensitive to data splits.
Detailed information on the training, validation, and test split, as well as the ten-fold split is given in the released dataset (see Data Availability).

\subsection*{AutoMatmainer training}

Automatminer is a machine learning pipeline that automatically featurizes the crystals and selects the appropriate features to train a set of machine learning algorithms \cite{dunn2020benchmarking}.
The best-performing algorithm is used as the final model.
For all the results reported in this work, the \verb|production| preset is adopted.
It was found that the gradient boost, random forest, and extra trees algorithms can all be selected as the best-performing model, depending on the target elastic property and the initialization of the parameters in the automatic pipeline.
For each target elastic property, the reported results are obtained by averaging over multiple runs, each with a different initialization.

\subsection*{Compliance tensor}

The compliance tensor $\bm S$ is a fourth-rank tensor defined from the inverse stress-strain relation $\bm \epsilon = \bm S \bm \sigma$, where $\bm\epsilon$ and $\bm\sigma$ are the second-rank strain tensor and stress tensor, respectively.
The compliance tensor in Voigt notation $\bm s$ can be obtained as the inverse of the elasticity tensor Voigt matrix,
\begin{equation}
    \bm s = \bm c^{-1},
\end{equation}
which is a 6 by 6 symmetric matrix.
The relationships between the components of the full compliance tensor $S_{ijkl}$ and the Vogit matrix $s_{\alpha\beta}$ are \cite{nye1985physical}:
\begin{equation}
    \begin{aligned}
        S_{ijkl}  & = s_{\alpha\beta},\, \text{when $\alpha$ and $\beta$ are 1, 2, or 3,}       \\
        2S_{ijkl} & = s_{\alpha\beta},\, \text{when either $\alpha$ or $\beta$ are 4, 5, or 6,} \\
        4S_{ijkl} & = s_{\alpha\beta},\, \text{when both $\alpha$ and $\beta$ are 4, 5, or 6.}  \\
    \end{aligned}
\end{equation}

\subsection*{Averaged elastic moduli of polycrystals}

Given the elasticity tensor of a single crystal, the averaged bulk, shear, and Young's moduli of polycrystalline materials can be computed using different average schemes.
The Voigt average assumes that the strain is the same in each grain, equal to the macroscopically applied strain \cite{voigt1910lehrbuch}.
The Voigt bulk modulus is
\begin{equation}
    K_V =  [(c_{11} + c_{22} + c_{33}) + 2(c_{12} + c_{23} + c_{31})]/9,
\end{equation}
and the Voigt shear modulus is
\begin{equation}
    G_V =  [
    (c_{11} + c_{22} + c_{33})
    -  (c_{12} + c_{23} + c_{31})
    + 3(c_{44} + c_{55} + c_{66}) ]
    / 15  .
\end{equation}
The Reuss average assumes that the stress is the same in each grain, equal to the macroscopically applied stress \cite{reuss1929berechnung}.
The Reuss bulk modulus is
\begin{equation}
    K_R =  1 / [(s_{11} + s_{22} + s_{33}) +  2(s_{12} + s_{23} + s_{31})] ,
\end{equation}
and Reuss the shear modulus is
\begin{equation}
    G_R =  15 /[
    4(s_{11} + s_{22} + s_{33})
    - 4(s_{12} + s_{23} + s_{31})
    + 3(s_{44} + s_{55} + s_{66})
    ].
\end{equation}
The Voigt and Reuss averages are the two extreme cases.
The Hill average takes their arithmetic mean and is considered the most accurate in a wide range of experimental conditions \cite{hill1952elastic}.
The Hill bulk modulus is
\begin{equation}
    K_H = (K_V + K_R)/2,
\end{equation}
and the Hill shear modulus is
\begin{equation}
    G_H = (G_V + G_R)/2.
\end{equation}
Given the bulk modulus and the shear modulus (from any of the Voigt, Reuss, and Hill schemes), Young's modulus can be computed as  \cite{anand2022introduction}
\begin{equation}
    E = 9KG/(3K + G).
\end{equation}
In this work, we report the bulk modulus $K=K_H$, shear modulus $G=G_H$, and Young's modulus $E = 9K_HG_H/(3K_H + G_H)$ from the Hill average scheme.

\subsection*{Scaled error}

The mean absolute error (MAE) and mean absolute deviation (MAD) are defined as
$\text{MAE} = \frac{1}{N_1} \sum_i^{N_1} |y_i - y_i^\text{pred}| $
and
$\text{MAD} = \frac{1}{N_2} \sum_i^{N_2} |(y_i - \bar y)|$,
in which $y_i$ is the reference value of data point $i$, $y_i^\text{pred}$ is the model prediction for the data point, and $\bar y$ is the average of all reference values.
The numbers $N_1$ and $N_2$ do not necessarily need to be the same.
This is the case in \fref{fig:error:scalar} and \tref{tab:mae:mad}, where $N_1$ is the number of crystals in a specific crystal system and $N_2$ is the total size of the test set.
The scaled error (SE) is then computed as $\text{SE} = \text{MAE} / \text{MAD}$.

\subsection*{Software implementation}
The \net model was implemented using the \verb|e3nn| package \cite{e3nnpaper} built on top of \verb|PyTorch| \cite{paszke2019pytorch},
and the training was performed using \verb|Pytorch-Lightning| \cite{lightning}.
The DFT calculations were performed using the \verb|atomate| workflow \cite{mathew2017atomate} and all crystal structure processing was performed using the Python Materials Genomics (\verb|pymatgen|) \cite{ong2013python}.
Directional Young's modulus was obtained using the \verb|elate| package \cite{gaillac2016elate}.

\section*{Data Availability}
The elasticity tensors used for model development,
the 100 new crystals with large maximum directional Young's modulus, and the elemental cubic metals are available at \url{https://doi.org/10.5281/zenodo.8190849}
The elasticity tensors are also available from the Materials Project database via the web interface at \url{https://materialsproject.org}
or the API at \url{https://api.materialsproject.org}.

\section*{Code Availability}
The \net model and training scripts are released as an open-source repository at \url{https://github.com/wengroup/matten}.

\section*{Acknowledgements}

The method development was supported by the National Science Foundation under Grant No.\ 2316667 and the startup funds from the Presidential Frontier Faculty Program at the University of Houston.
Support for software and data infrastructure development was provided by the U.S.\ Department of Energy, Office of Science, Office of Basic Energy Sciences, Materials Sciences and Engineering Division under contract No.\ DE-AC02-05-CH11231 (Materials Project program KC23MP).
This work used computational resources provided by
the Research Computing Data Core at the University of Houston,
the National Energy Research Scientific Computing Center (NERSC), a U.S.\ Department of Energy Office of Science User Facility operated under contract No.\ DE-AC02-05CH11231,
and the Lawrencium computational cluster resource provided by the IT Division at the Lawrence Berkeley National Laboratory (Supported by the Director, Office of Science, Office of Basic Energy Sciences, of the U.S.\ Department of Energy under contract No.\ DE-AC02-05CH11231).

\section*{Author Contributions}

Conceptualization, investigation, software, visualization, and writing - original draft: M.W.;
data curation and formal analysis: M.W., M.K.H., J.M.M., and P.H.;
writing - review \& editing: M.W., M.K.H., J.M.M., P.H., and K.A.P.;
project administration: M.W.;
funding acquisition: M.W. and K.A.P.

\section*{Conflicts of Interests}

There are no conflicts to declare.

\bibliographystyle{unsrtnat}

\end{document}


\maketitle

\section*{The ten symmetry classes of elasticity tensors}

\begin{figure}
\centering
    \includegraphics[width=0.8\columnwidth]{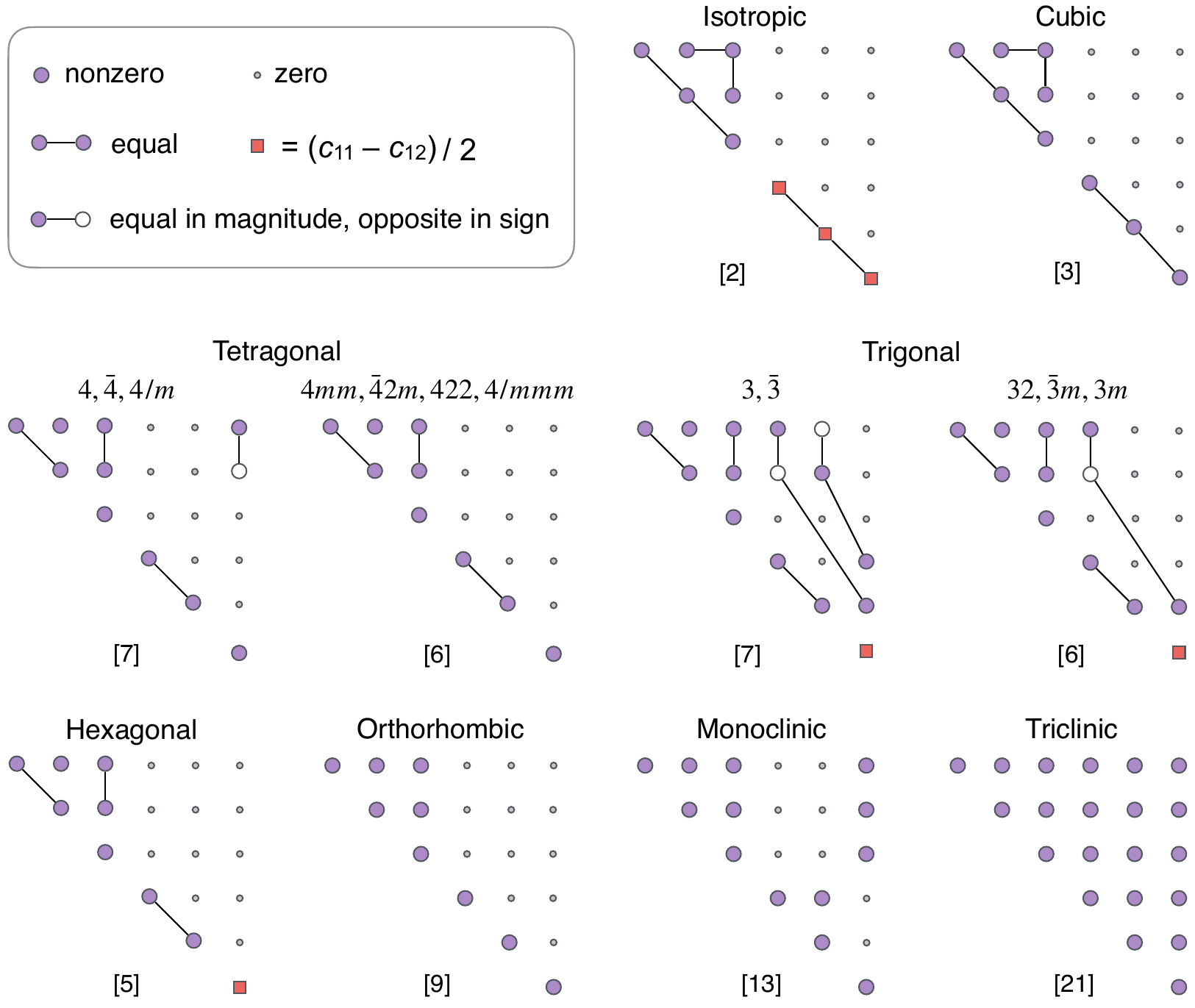}
    \caption{Symmetry classes and independent components of the stiffness tensor according to the crystallographic Laue group by Wallace \cite{wallace1972thermodynamics}. 
    There are 11 classes from Laue group, with two for the monoclinic system depending on the orientation. 
    Here we only depict the one with the standard orientation. 
    See \onlinecite{nye1985physical} for the other case.
    The conclusion is incorrect for tetragonal and trigonal crystal systems---for each of them, there are two cases with 6 and 7 independent components. 
    The value in the square brackets is the number of independent components for the corresponding crystal system. All matrices are symmetric about the leading diagonal, with the lower left part omitted in the depiction. 
    }
    \label{fig:10:class}
\end{figure}

Early approaches take inspiration from crystallography. 
Out of the 32 distinct crystallographic point groups, only 11 are centrosymmetric (meaning the point group contains an inversion center as one of its symmetry elements), each forming a unique diffraction pattern. 
The diffraction patterns of other noncentrosymmetric crystals each is the same as one of the 11 centrosymmetric crystals. 
Based on the diffraction patterns, the 32 distinct point groups can be categorized into 11 classes, called the Laue groups \cite{tadmor2012continuum}.
According to the Laue groups, Wallace \cite{wallace1972thermodynamics} classifies the elasticity tensors into 12 classes (the additional 1 being the isotropic class that does not apply for single crystals), and they reduce to 10 classes considering the number of independent components (\fref{fig:10:class}). 
The results are widely cited, including the classical book on the subject by Nye \cite{nye1985physical} and many recent papers \cite{singh2021mechelastic,li2022elast,ran2023velas}.
This crystallographic approach seems reasonable; however, the conclusions are incorrect.
The tetragonal and trigonal systems are each divided into two symmetry classes, 
but the distinctions can be eliminated by a different choice of the coordinate system \cite{fedorov1968theory,sutcliffe1992spectral}. 
Then each of the tetragonal and trigonal systems will have 6 independent components.

\section*{Harmonic decomposition of the elasticity tensor}

In the harmonic decomposition, the elasticity tensor can be written as 
\begin{equation}
    \bm C = h_1 (\lambda)  + h_2(\eta) + h_3 (\bm A) + h_4 (\bm B) +  h_5(\bm H).  
\end{equation}

The appropriate values for each of the term is as follows:
\begin{equation*}
    \lambda  =  [2C_{ppmm} - C_{pmpm}] / 15,
\end{equation*}
\begin{equation*}
    \eta =  [-3C_{ppmm} - C_{pmpm} ] /90,
\end{equation*}
\begin{equation*}
   A_{ij}  = [ 15C_{ijmm} - 12C_{imjm} -5\delta_{ij}C_{ppmm}      +4\delta_{ij}C_{pmpm} ]/ 21 ,
\end{equation*}
\begin{equation*}
   B_{ij}  = [-6C_{ijmm} + 9C_{imjm} +2\delta_{ij}C_{ppmm}      -3\delta_{ij}C_{pmpm} ] / 21 ,
\end{equation*}
\begin{equation*}
\begin{aligned}
H_{ijkl} 
=& (C_{ijkl} +  C_{iklj} + C_{iljk})/3 \\
 &- [(C_{ijmm} +  2C_{imjm} )\delta_{kl}
 + (C_{klmm} +  2C_{kmlm} )\delta_{ij} \\
 &+ (C_{ikmm} +  2C_{imkm} )\delta_{jl}
 + (C_{jlmm} +  2C_{jmlm} )\delta_{ik}  \\
 &+ (C_{ilmm} +  2C_{imlm} )\delta_{jk} 
 + (C_{jkmm} +  2C_{jmkm} )\delta_{il} ] / 21 \\
 &+ (C_{ppmm} +  2C_{pmpm} ) 
 (\delta_{ij}\delta_{kl} 
 +\delta_{ik}\delta_{jl}
 +\delta_{il}\delta_{jk}
 ) /105 ,
\end{aligned}
\end{equation*}
and 
\begin{equation*}
h_1 (\lambda) = \delta_{ij}\delta_{kl}\lambda,
\end{equation*}
\begin{equation*}
h_2 (\eta) = (\delta_{ik}\delta_{jl} + \delta_{il}\delta_{jk} )\eta ,
\end{equation*}
\begin{equation*}
h_3 (\bm A) = \delta_{ij}A_{kl} + \delta_{kl}A_{ij} , 
\end{equation*}
\begin{equation*}
h_4 (\bm B) 
= \delta_{ik}B_{jl} + \delta_{jl}B_{ik} 
+ \delta_{il}B_{jk} + \delta_{jk}B_{il} ,
\end{equation*}
\begin{equation*}
h_5 (\bm H)  = H_{ijkl},
\end{equation*}
where $\delta_{ij}$ is the Kronecker delta.

This decomposition follows \onlinecite{forte1996symmetry}, and as mentioned there that ``... other forms of harmonic decomposition are possible: It suffices to use invertible linear combinations of $\bm A$ and $\bm B$ and, analogously, invertible linear combinations of $\lambda$ and $\eta$.''
See \onlinecite{backus1970geometrical,itin2020irreducible} for such examples.
Nevertheless, harmonic decomposition is unique to linear combinations. 

This harmonic decomposition can be easily carried out with the \verb|e3nn| package \cite{e3nnpaper}. It can deal with any tensor of any symmetry, and below is a code snippet to
 obtain the irreducible representations of the elasticity tensor from the harmonic decomposition.

\begin{python}
>>> from e3nn import o3, io

>>> tp = o3.ReducedTensorProducts("ijkl=jikl=ijlk=klij", i="1o")
>>> tp.irreps_out
2x0e+2x2e+1x4e

# Alternatively
>>> ct = io.CartesianTensor("ijkl=jikl=ijlk=klij")
>>> ct
2x0e+2x2e+1x4e
\end{python}

The \verb|2x0e|, \verb|2x2e|, and \verb|4e| represent the two isotropic terms, the two deviatoric terms, and the harmonic term, respectively.

\section*{Proof of \net satisfying material symmetry}

The \net model $\bm C = f(x)$ is equivariant to $SO(3)$ transformations, satisfying 
\begin{equation} \label{eq:equi}
   D_y(g) f(x) = f( D_x(g) x) .
\end{equation}
This comes from the fact that each layer of \net is equivariant, and the composition of such layers is also equivariant. 
We refer to \onlinecite{thomas2018tensor} for proof of the equivariance of the layers.
The representation $D_x(g)$ in the space of crystal structures can be written as  $D_x(g) = R_{ip}$, and the representation $D_y(g)$ in the space of stiffness tensors can be written as  $D_g(g) = R_{ip}R_{jq}R_{kr}R_{ls}$, where $R \in SO(3)$ is a rotation matrix.

Let $Q \in P$, where $P$ denotes the set of rotations in the point group of a crystal, we will have $P \subset SO(3)$.
Therefore, for $R = Q$, \eref{eq:equi} is satisfied, i.e.,
\begin{equation}
   Q_{ip}Q_{jq}Q_{kr}Q_{ls} C_{prqs} = f(Qx)
\end{equation}
Owning to material symmetry, we have $Qx = x$, that is, the crystal structure $x$ is indistinguishable before and after the transformation.  
Thus, $f(Qx) = f(x)$. 
Plugging it into \eref{eq:equi}, we have 
\begin{equation}
   Q_{ip}Q_{jq}Q_{kr}Q_{ls} C_{prqs} = f(x) = C_{ijkl},
\end{equation}
which is Eq.(1) in the main text.
Once this is satisfied, the material symmetry will be reflected in the stiffness tensor as discussed in the main text and proved in \onlinecite{forte1996symmetry}.

\section*{Dataset statistics}

\begin{figure}[H]
    \centering
    \includegraphics[width=0.55\columnwidth]{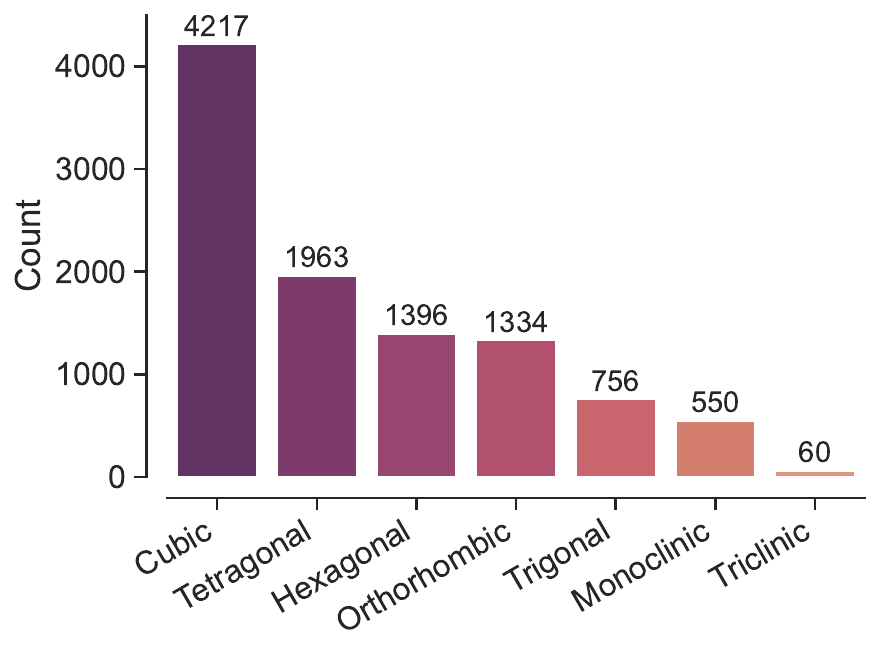}
    \caption{Histogram of the dataset by crystal system. The dataset consists of a total number of 10276 elasticity tensors for inorganic crystals.  
    }
    \label{fig:data:count}
\end{figure}

\begin{figure}[H]
    \centering
    \includegraphics[width=0.55\columnwidth]{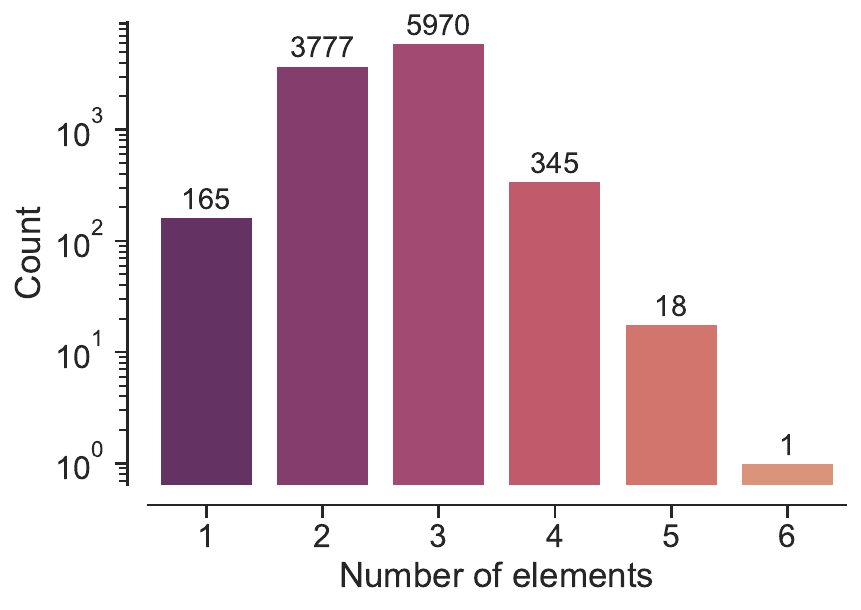}
    \caption{Histogram of the dataset by the number of chemical elements in the crystals.   
    }
    \label{fig:data:count:elements}
\end{figure}

\begin{figure}[H]
    \centering
    \includegraphics[width=0.95\columnwidth]{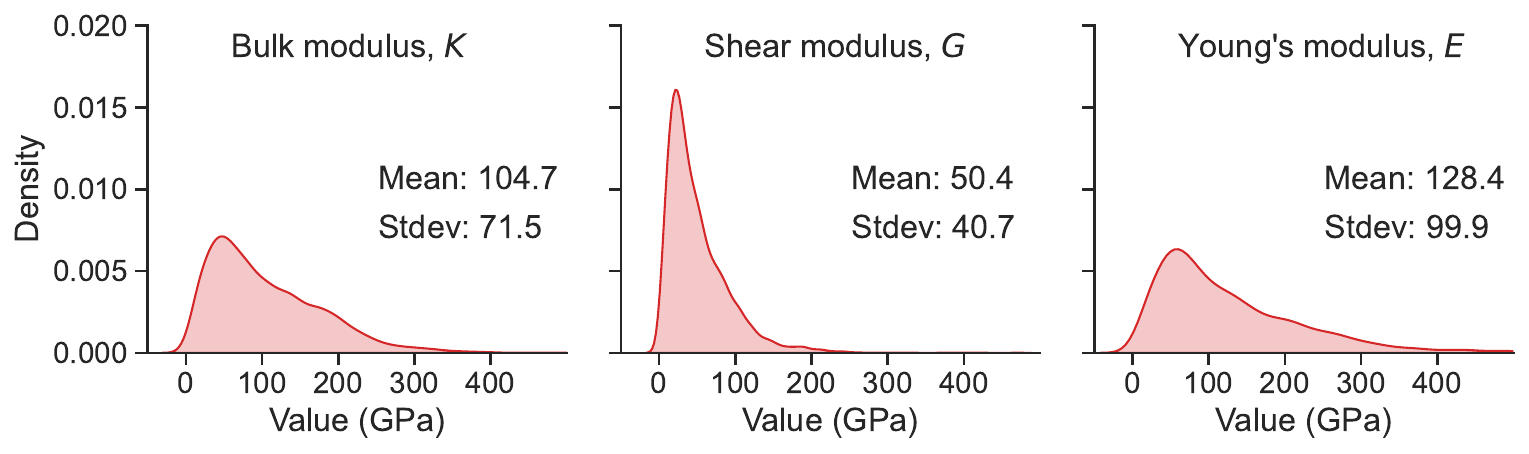}
    \caption{Distribution of the bulk, shear, and Young's moduli in the dataset.
    }
    \label{fig:k:g:e:density}
\end{figure}

\begin{figure}[H]
    \centering
    \includegraphics[width=0.95\columnwidth]{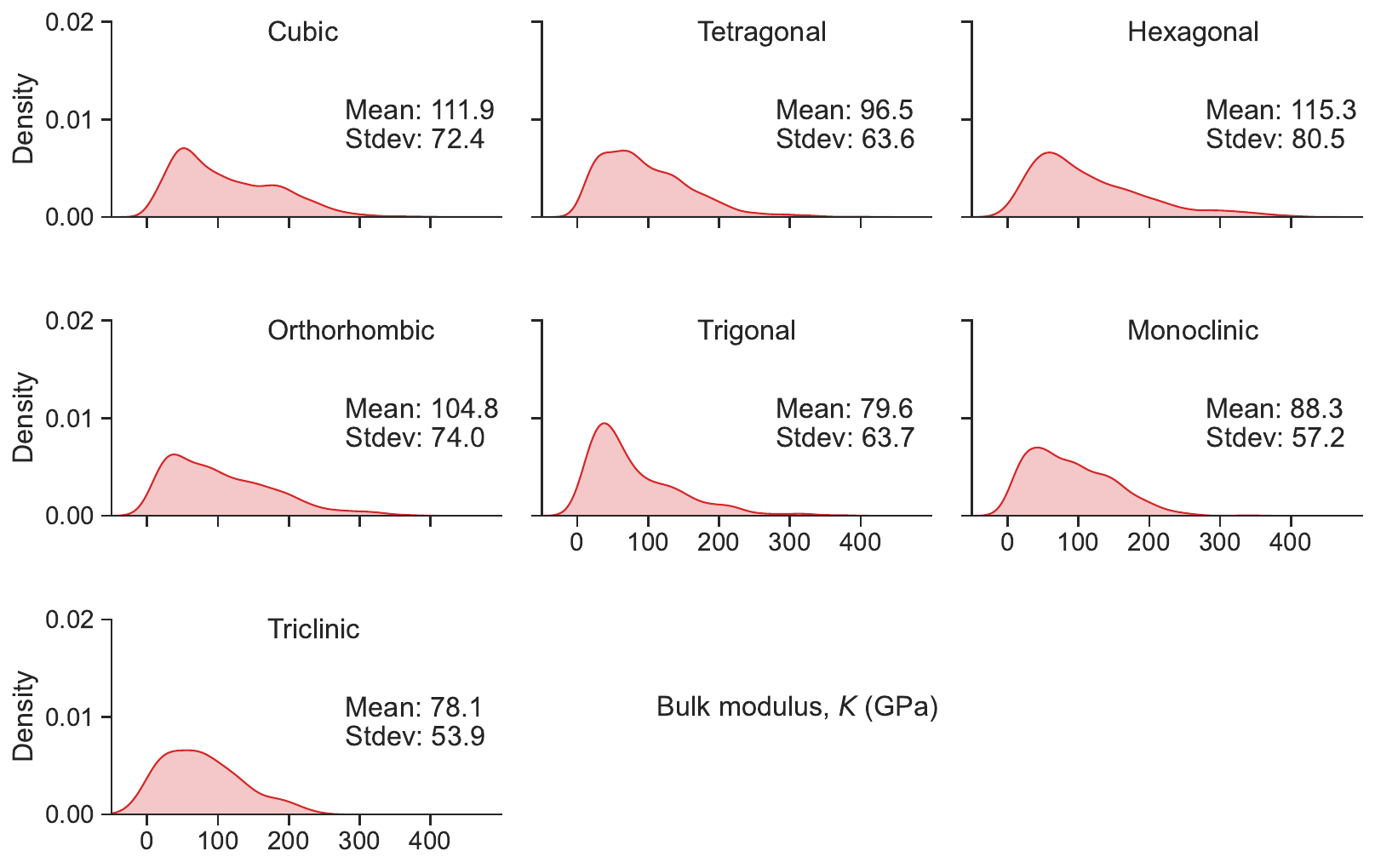}
    \caption{Distribution of bulk modulus in the dataset by crystal system.
    }
\end{figure}

\begin{figure}[H]
    \centering
    \includegraphics[width=0.95\columnwidth]{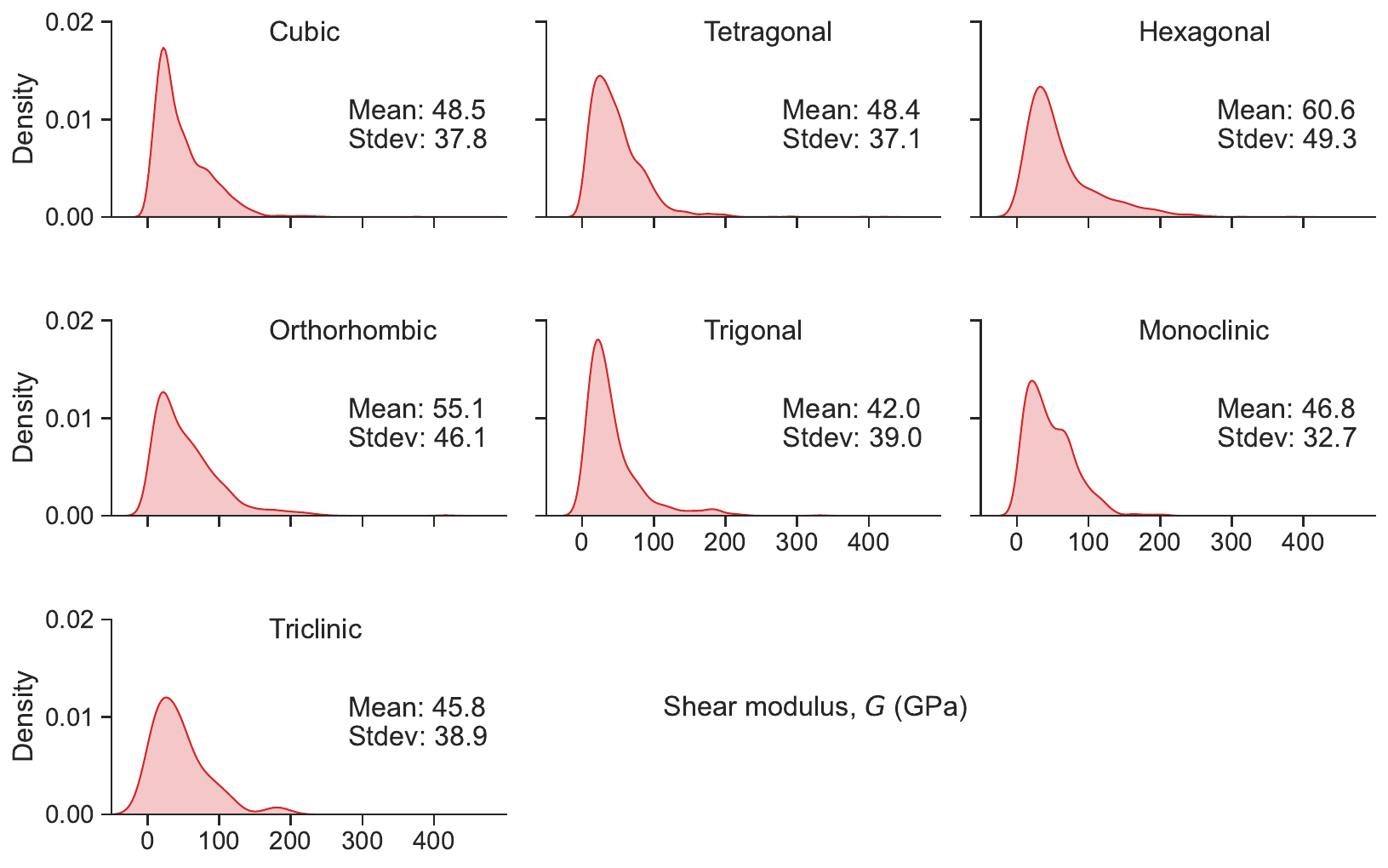}
    \caption{Distribution of shear modulus in the dataset by crystal system.
    }
\end{figure}

\begin{figure}[H]
    \centering
    \includegraphics[width=0.95\columnwidth]{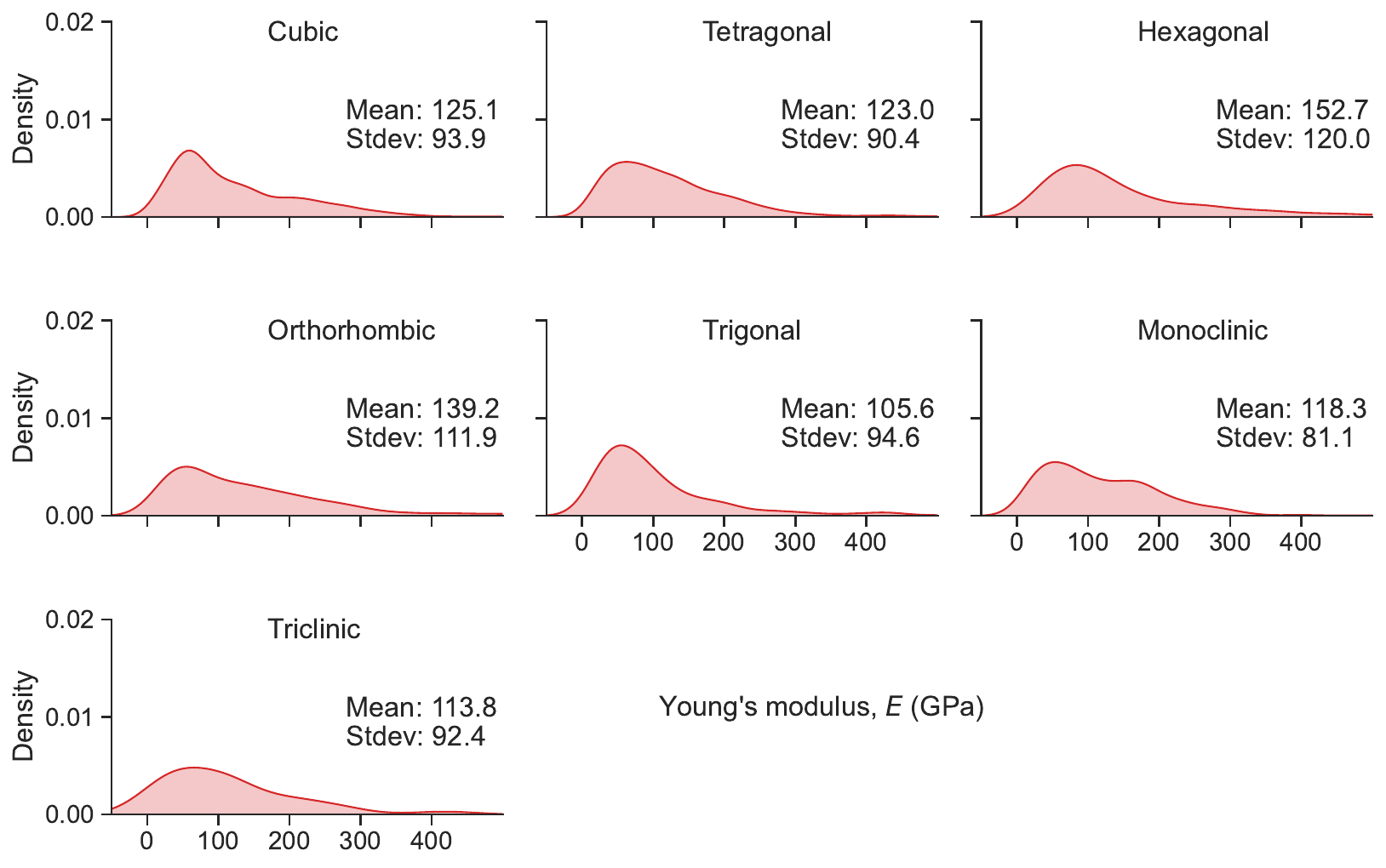}
    \caption{Distribution of Young's modulus in the dataset by crystal system.
    }
\end{figure}

\section*{Error in strain caused by that in Young's modulus}

\begin{figure}[H]
    \centering
    \includegraphics[width=0.3\columnwidth]{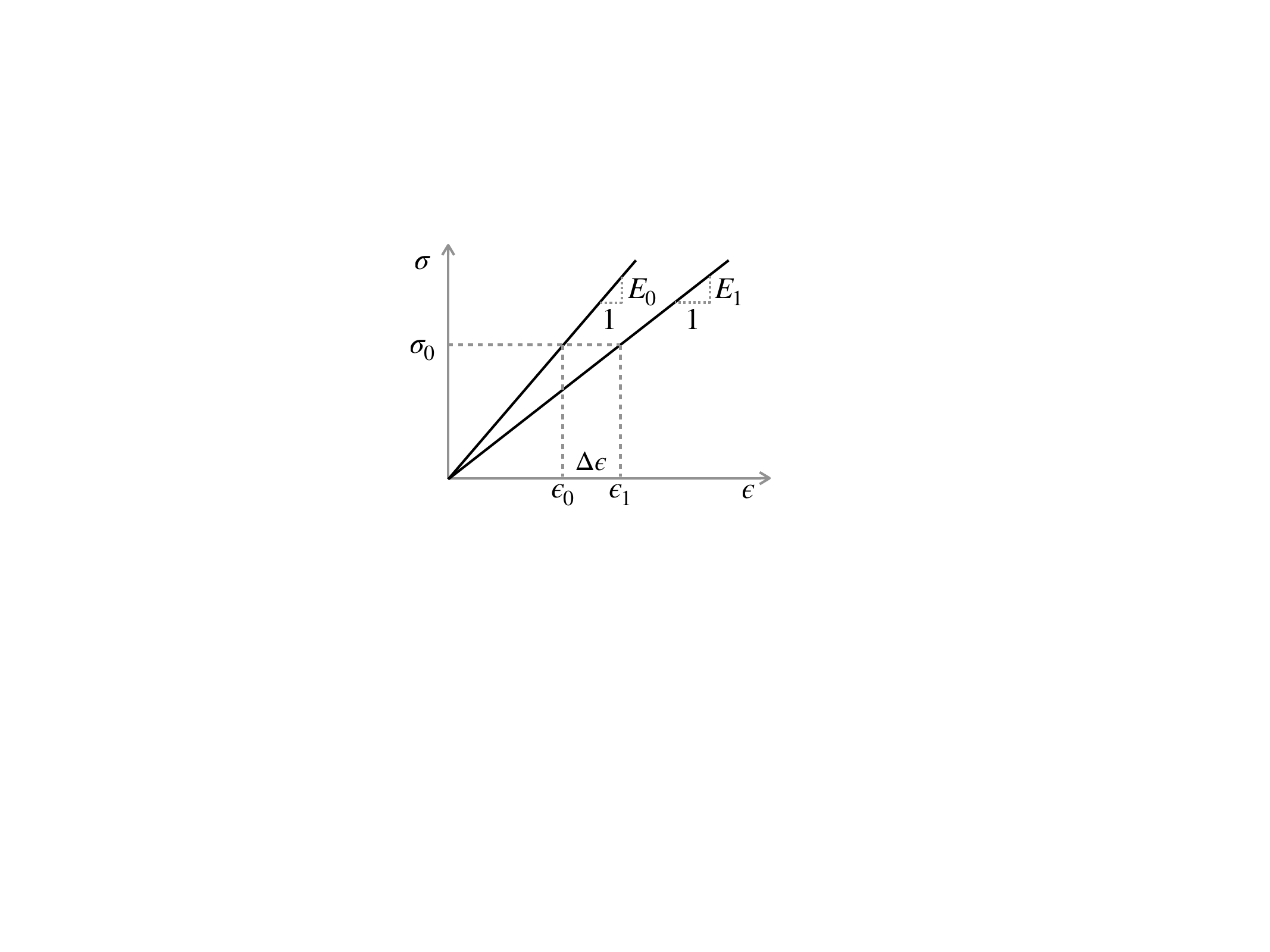}
    \label{fig:error:strain}
\end{figure}

We consider the strain change due to error in Young's modulus under the same stress $\sigma_0$.
Let $E_0 = 128.4$~GPa (mean of DFT reference values) and $E_1 = E_0 - \Delta E$, where $\Delta E = 20.59$~GPa is the mean absolute error (MAE) of \net predictions. 
We have 
\begin{equation}
\begin{aligned}
\sigma_0 &= E_0\epsilon_0 \\
\sigma_0 &= E_1\epsilon_1 = (E_0 - \Delta E) (\epsilon_0 + \Delta \epsilon) .
\end{aligned}
\label{eq:strain:error}
\end{equation}
Solve \eref{eq:strain:error}, we have 
\begin{equation}
    \Delta \epsilon = \left[\frac{E_0}{E_0 - \Delta E}  - 1 \right]\epsilon_0 
     = 19\% \ \epsilon_0
\end{equation}

\section*{Test errors}

\begin{figure}[H]
    \centering
    \includegraphics[width=0.6\columnwidth]{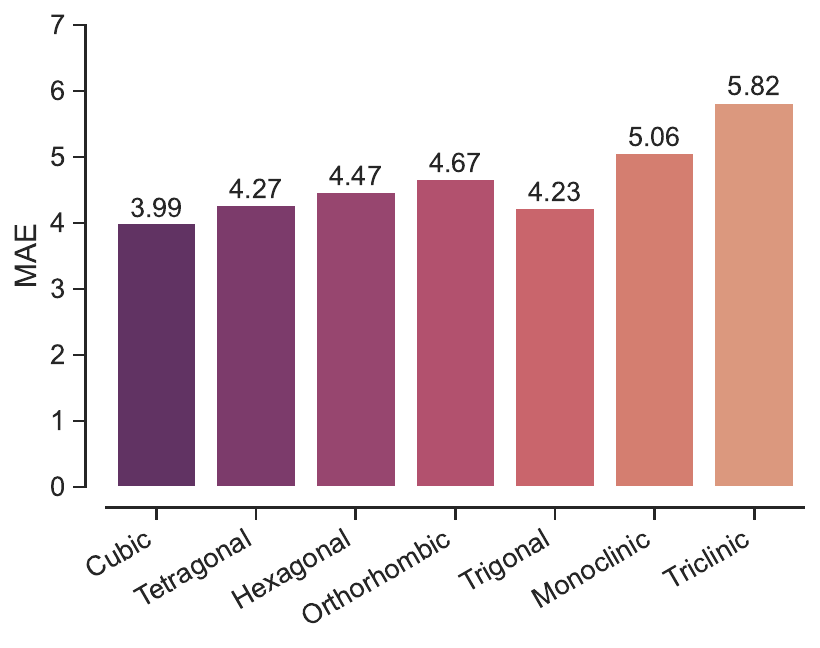}
    \caption{Mean absolute error (MAE) of the elasticity tensor by crystal system. The MAE is computed using the predicted and reference Voigt matrix of the elasticity tensor.
    }
    \label{fig:mae:voigt}
\end{figure}

\begin{figure}[H]
    \centering
    \includegraphics[width=0.6\columnwidth]{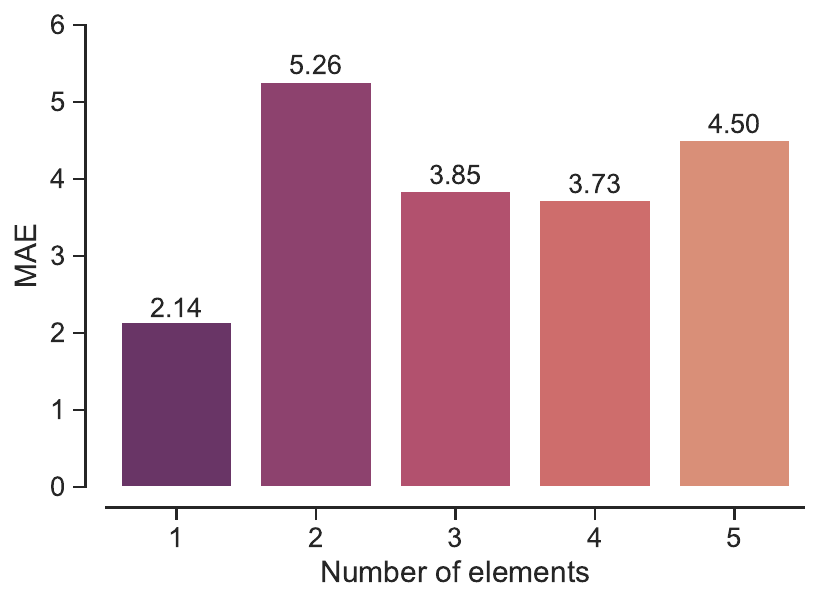}
    \caption{Mean absolute error (MAE) of the elasticity tensor by the number of chemical elements in the crystals. The MAE is computed using the predicted and reference Voigt matrix of the elasticity tensor.
    }
\end{figure}

\begin{figure}[H]
    \centering
    \includegraphics[width=0.6\columnwidth]{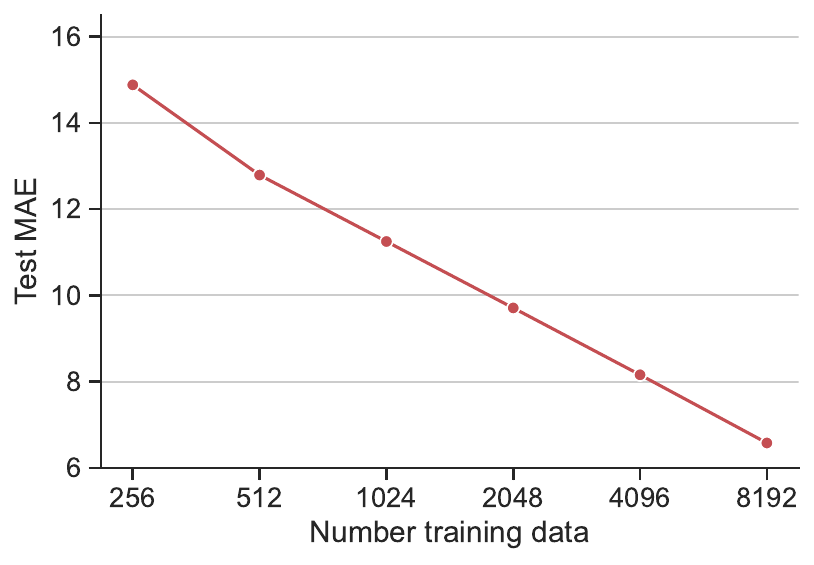}
    \caption{Learning curve of the \net model. The MAE is obtained using the predicted and reference elasticity tensors in Voigt notation.
    The MAE is on the test set, and the number of training data is sampled from the training set.
    }
    \label{fig:learning:curve}
\end{figure}

\begin{figure}[H]
    \centering
    \includegraphics[width=0.6\columnwidth]{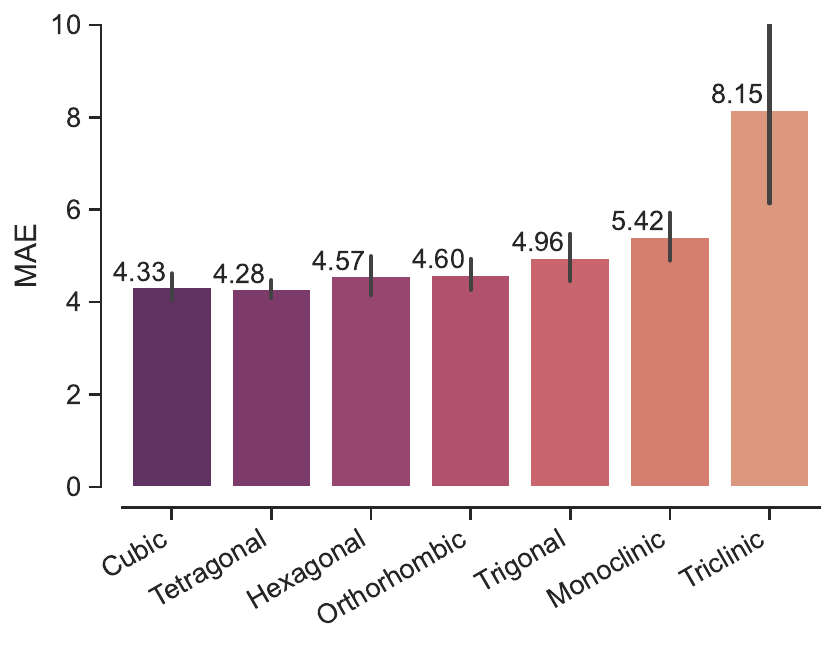}
    \caption{Mean absolute error (MAE) of the elasticity tensor in Voigt matrix computed from ten-fold cross validation.
    Compare with \fref{fig:mae:voigt}, where the 9th and 10th fold of the data are used as the validation and test set, respectively.
    }
\end{figure}

\begin{figure}[H]
    \centering
    \includegraphics[width=1.0\columnwidth]{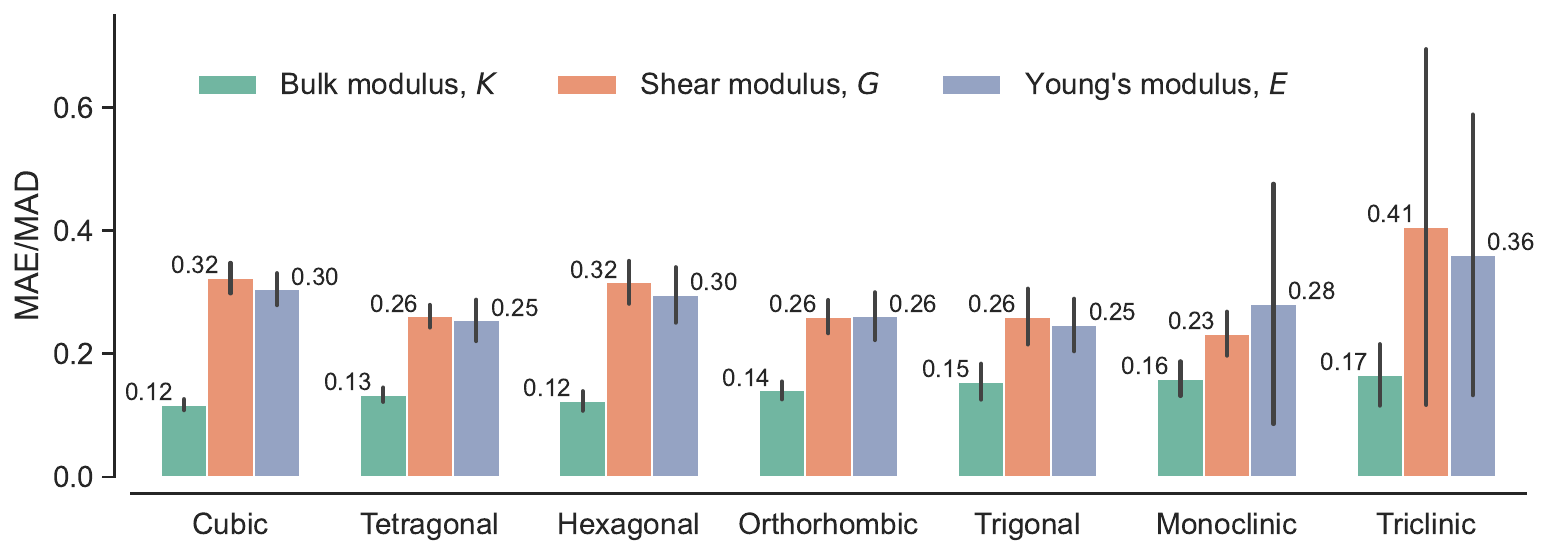}
    \caption{Ten-fold cross validation scaled error (MAE/MAD) on the bulk, shear, and Young's moduli. Compare with Fig.~3d in the main text.}
\end{figure}

\section*{Training on tensor components}

\begin{figure}[H]
    \centering
    \includegraphics[width=0.5\columnwidth]{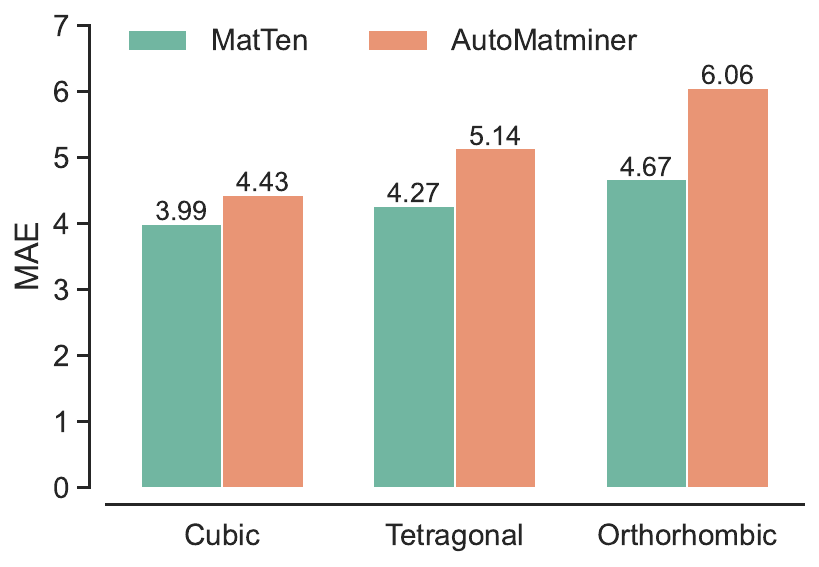}
    \caption{Mean absolute error (MAE) of the elasticity tensor in Voigt matrix.
    Multiple AutoMatminer models are trained for each crystal system, each model predicting a separate non-zero component of the Voigt matrix in Fig.~1 in the main text. 
    For example, for the orthorhombic crystal system, nine AutoMatminer models are trained.}
    \label{fig:tensor:components}
\end{figure}

It is possible to predict the full elasticity tensor by separately modeling its non-zero independent components.
Because each crystal system has a different number of non-zero components (Fig.~1 in the main text), this approach requires the treatment of each crystal system separately.
To check how this approach works, we consider the cubic, tetragonal, and orthorhombic crystal systems. 
For each of them, we select the corresponding crystals in the training, validation, and test sets, and then train multiple AutomMatminer models, each with one non-zero component of the full tensor as the target.
The mean absolute error (MAE) is shown \fref{fig:tensor:components}; also plotted are the \net results for comparison. 

For the ``training tensor components'' approach, the performance deteriorates quickly with the tensor complexity, i.e., the number of independent components in the tensor, increasing from cubic to tetragonal, and to orthorhombic. 
In contrast, the error by \net only slightly increases with increased tensor complexity, demonstrating the advantage of the united \net approach. 
\net automatically handles all symmetry requirements and thus allows the training using all data, irrespective of the crystal systems.  
This contributes to the improved performance of \net.

\section*{Additional results on isotropic properties}

\begin{table}[H]
\small
\caption{Prediction of the bulk modulus $K$, shear modulus $G$, and Young's modulus $E$ in logarithmic space.
$K$, $G$, and $E$ are in the units of GPa.
The results for \net are calculated from a single model, while a separate AutoMatminer model is trained for each property.
The value in a pair of parentheses is the standard deviation from an ensemble of five models trained with different initialization.
MAE: mean absolute error; MAD: mean absolute deviation.}
\label{tab:k:g:log}
\centering
\begin{tabular}{@{\extracolsep{5pt}}ccccccc}
\hline
  &\multicolumn{2}{c}{$\log_{10}(K)$}  
  &\multicolumn{2}{c}{$\log_{10}(G)$}  
  &\multicolumn{2}{c}{$\log_{10}(E)$}   \\
  \cline{2-3} \cline{4-5} \cline{6-7}
  & MAE  & MAE/MAD & MAE  & MAE/MAD  & MAE & MAE/MAD \\
  
\hline
MatTen        & 0.046 (0.002)  &0.166 (0.006)   & 0.094 (0.002) & 0.331 (0.010)  & 0.087 (0.002) & 0.309 (0.018) \\
AutoMatminer  & 0.050 (0.002)  & 0.187 (0.009)  & 0.090 (0.002) & 0.307 (0.006)  & 0.086 (0.002)  &0.301 (0.009)\\
 \hline
\end{tabular}
\end{table}

\section*{Failure analysis}
We checked the positive definiteness of the predicted elasticity tensors for the crystal in the test set.
The 25 cases with at least one negative eigenvalues are listed in \tref{tab:failure}.
For the cubic, tetragonal, and orthorhombic crystals, the failure happens all because of the incorrect prediction of the relative magnitude of the diagonal component and off-diagonal components. 
For example, for the orthorhombic \ce{Na4C4S4N4} crystal (mp-6633), the DFT elasticity tensor is: 
\[
\begin{bmatrix}
46.7 & 18.1  & 12.2 & 0.0 & 0.0 & 0.0 \\
18.1 & 30.8  & 10.3 & 0.0 & 0.0 & 0.0 \\
12.2 & 10.3 & 22.0 & 0.0 & 0.0 & 0.0 \\
0.0  & 0.0  & 0.0  & 7.4 & 0.0 & 0.0 \\
0.0  & 0.0  & 0.0  & 0.0 & 8.5 & 0.0 \\
0.0  & 0.0  & 0.0  & 0.0 & 0.0 & 10.2
\end{bmatrix}, 
\]
while the model predicted is:
\[
   \begin{bmatrix}
11.4 & 19.7 &  9.9 & 0.0 & 0.0 & 0.0 \\
19.7 & 24.2 &  3.2 & 0.0 & 0.0 & 0.0 \\
 9.9 &  3.2 & 18.5 & 0.0 & 0.0 & 0.0 \\
 0.0 &  0.0 &  0.0 &11.8 & 0.0 & 0.0 \\
 0.0 &  0.0 &  0.0 & 0.0 &10.3 & 0.0 \\
 0.0 &  0.0 &  0.0 & 0.0 & 0.0 &11.0
\end{bmatrix} .
\]
The predicted $c_{11}$ is substantially smaller than the DFT value.
For the more complex (in terms of the number of independent components) trigonal crystals, we did not observe any pattern.
Nor for the two monoclinic crystals.

\begin{table}[H]
\small
\caption{Number of crystals with negative eigenvalues by crystal system.}
\label{tab:failure}
\centering
\begin{tabular}{ccccccc}
\hline 
Cubic &  Tetragonal & Hexagonal & Orthorhombic & Trigonal & Monoclinic &Triclinic \\
\hline
 7 &   7 &  0 & 4  &  5  & 2  & 0   \\
\hline
\end{tabular}
\end{table}

\section*{Directional Young's modulus}

Here we prove that, for cubic crystals,
\begin{equation}\label{eq:max:E:dir:less}
\text{if}\, S_{1111} - S_{1122} - 2S_{2323} < 0, 
E_\text{d}^\text{max}\, \text{is along}\, \langle100\rangle\, \text{and}\,
E_\text{d}^\text{max}\, \text{is along}\, \langle111\rangle,
\end{equation}
otherwise,
\begin{equation}\label{eq:max:E:dir:greater}
\text{if}\, S_{1111} - S_{1122} - 2S_{2323} > 0, 
E_\text{d}^\text{max}\, \text{is along}\, \langle111\rangle\, \text{and}\,
E_\text{d}^\text{max}\, \text{is along}\, \langle100\rangle.
\end{equation}
and 
\begin{equation}\label{eq:max:E:dir:equal}
\text{if}\, S_{1111} - S_{1122} - 2S_{2323} = 0, \,
\text{the materials is isotropic regarding Young's modulus}.
\end{equation}

The inverse of the directional Young's modulus is 
\begin{equation} \label{eq:E:d:si}
       E_\text{d} ({\bm n})^{-1}  =n_i n_j n_k n_l S_{ijlk},
\end{equation}
where $S_{ijkl}$ is the compliance tensor and $\bm n$ is an unit direction vector.
For a cubic crystal, the 21 non-zero components can be classified into three groups \cite{nye1985physical}: 
\begin{itemize}
    \item $S_{1111} = S_{2222} = S_{3333}$ 
    \item $S_{1122} = S_{2211} = S_{2233} = S_{3322} = S_{3311} = S_{1133}$ 
    \item $  S_{2323} = S_{2332} = S_{3223} = S_{3232} 
           = S_{1212} = S_{1221} = S_{2112} = S_{2121} 
           = S_{1313} = S_{1331} = S_{3113} = S_{3131} $. 
\end{itemize}
Substituting these into \eref{eq:E:d:si}, we have, 
\begin{equation} \label{eq:E:d:cubic}
\begin{aligned}
 E_\text{d} ({\bm n}) ^{-1}
 &= S_{1111}(n_1^4 + n_2^4 + n_3^4) 
 + 2S_{1122}(n_1^2n_2^2 + n_2^2n_3^2 + n_3^2n_1^2) 
 + 4S_{2323}(n_1^2n_2^2 + n_2^2n_3^2 + n_3^2n_1^2)   \\
 &=  S_{1111} [1 - 2 (n_1^2n_2^2 + n_2^2n_3^2 + n_3^2n_1^2)]
 + 2S_{1122}(n_1^2n_2^2 + n_2^2n_3^2 + n_3^2n_1^2) 
 + 4S_{2323}(n_1^2n_2^2 + n_2^2n_3^2 + n_3^2n_1^2)  \\
 & = S_{1111} - 2(S_{1111} - S_{1122} - 2S_{2323})  (n_1^2n_2^2 + n_2^2n_3^2 + n_3^2n_1^2)  \\
 & = S_{1111} - 2(S_{1111} - S_{1122} - 2S_{2323})  f. 
\end{aligned}
\end{equation}
In the second equality, we have used 
$n_1^4 + n_2^4 + n_3^4
= (n_1^2 + n_2^2 + n_3^2)^2 - 2 (n_1^2n_2^2 + n_2^2n_3^2 + n_3^2n_1^2)
= 1 - 2 (n_1^2n_2^2 + n_2^2n_3^2 + n_3^2n_1^2)
$, in which $(n_1^2 + n_2^2 + n_3^2)^2 = 1$, because $\bm n$ is a unit vector.
In the last equality, we have defined  $f := n_1^2n_2^2 + n_2^2n_3^2 + n_3^2n_1^2$. 

From \eref{eq:E:d:cubic}, it seen that \eref{eq:max:E:dir:equal} is valid.

In fact, $f$ has its maximum value of $1/3$ along the $\langle 111 \rangle$ directions, and the minimum of $f$ is 0 along the $\langle 100 \rangle$ directions (derived below).
As a result, \eref{eq:max:E:dir:less} and \eref{eq:max:E:dir:greater} are valid.
(Note that \eref{eq:E:d:cubic} gives the inverse of the directional Young's modulus.)

Below, we show that the maximum of $f$ is $1/3$ along the $\langle 111 \rangle$ directions, and the minimum of $f$ is 0 along the $\langle 100 \rangle$ directions.

Let $n_1^2 = a, n_2^2 = b $ and $n_3^2 = c$, we have $a + b + c = 1$ because $\bm n$ is a unit vector.
Thus, 
\begin{equation} \label{eq:f}
f   
= ab + bc + ca  
= ab + c(b + a)  
= ab + (1-a - b)(b + a)  
= a + b - ab - a^2 - b^2  .
\end{equation}
Let 
\begin{equation}
\begin{aligned}
\frac{\partial f}{\partial a} & = 1 - b - 2a = 0 \\
\frac{\partial f}{\partial b} & = 1 - a - 2b = 0 , 
\end{aligned}
\end{equation}
and solve the equations, we have $a = b = c = 1/3$, i.e.\  $n_1^2 = n_2^2 = n_3^3 = 1/3$, 
At these values, $f=1/3$ and we can verify that it is a maximum.
This also suggests $\bm n$ is along the $\langle 111 \rangle$ family of directions. 

The other extreme values of $f$ are located at the boundaries of $a$ (or $b$ or $c$). Since $n_1$ is a component of the unit vector, then $n_1 \in [-1, 1]$, i.e.\  $a \in [0, 1]$. So, the extreme value is obtained when
\begin{itemize}
    \item  $n_1 = 0$, $n_2 = \pm 1$, $n_3 = 0$
    \item  $n_1 = 0$, $n_2 = 0$, $n_3 = \pm 1$
    \item  $n_1 = \pm 1$, $n_2 = 0$, $n_3 = 0$.
\end{itemize} 
These are the $\langle 100 \rangle$ directions, at which the minimum is $f=0$.

\begin{figure}[H]
    \centering
    \includegraphics[width=1.0\columnwidth]{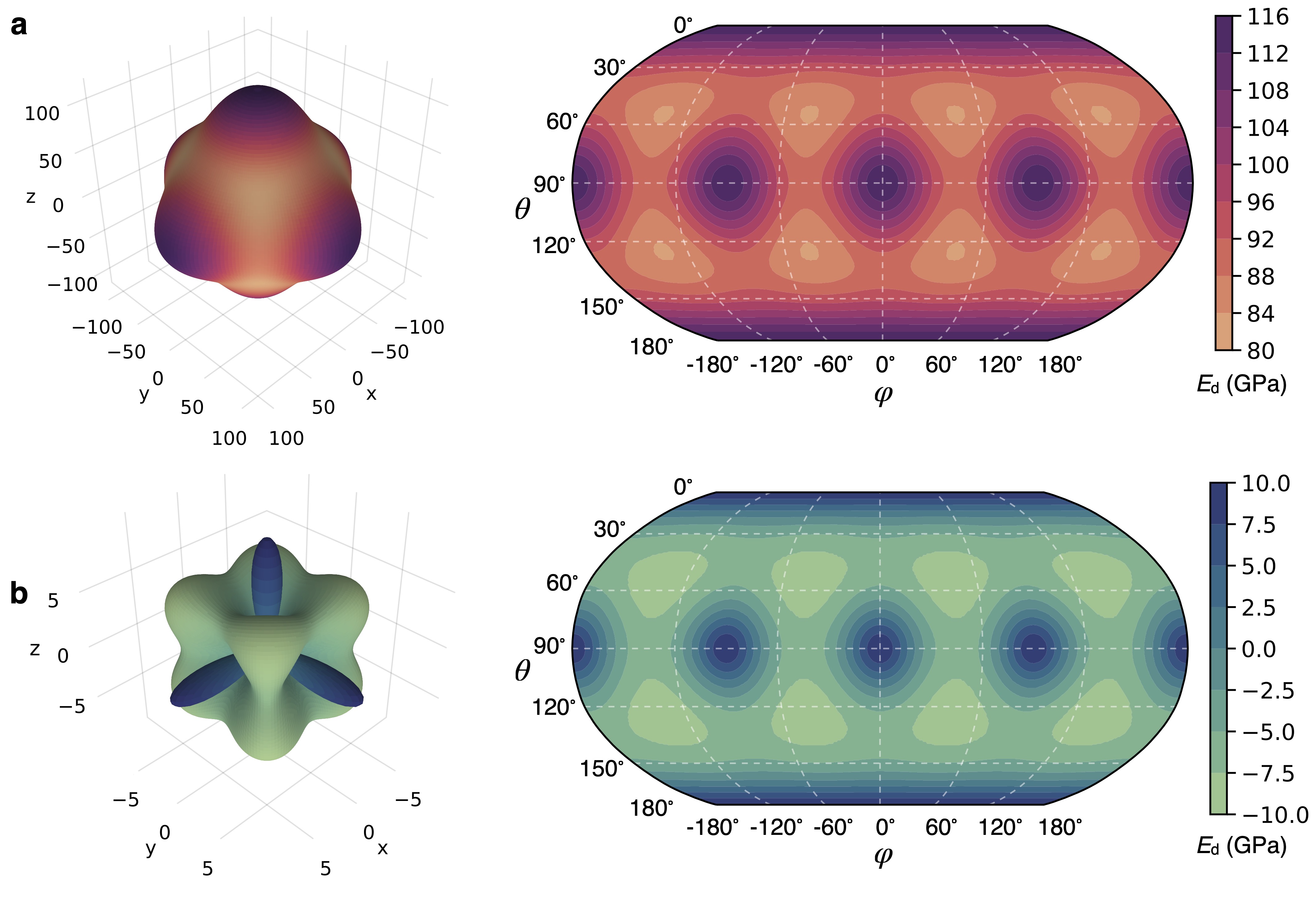}
    \caption{Directional Young's modulus $E_\text{d}$ for CaS. (a) DFT reference values. (b) Prediction error between \net prediction and DFT reference values.}
\end{figure}

\begin{figure}[H]
    \centering
    \includegraphics[width=0.6\columnwidth]{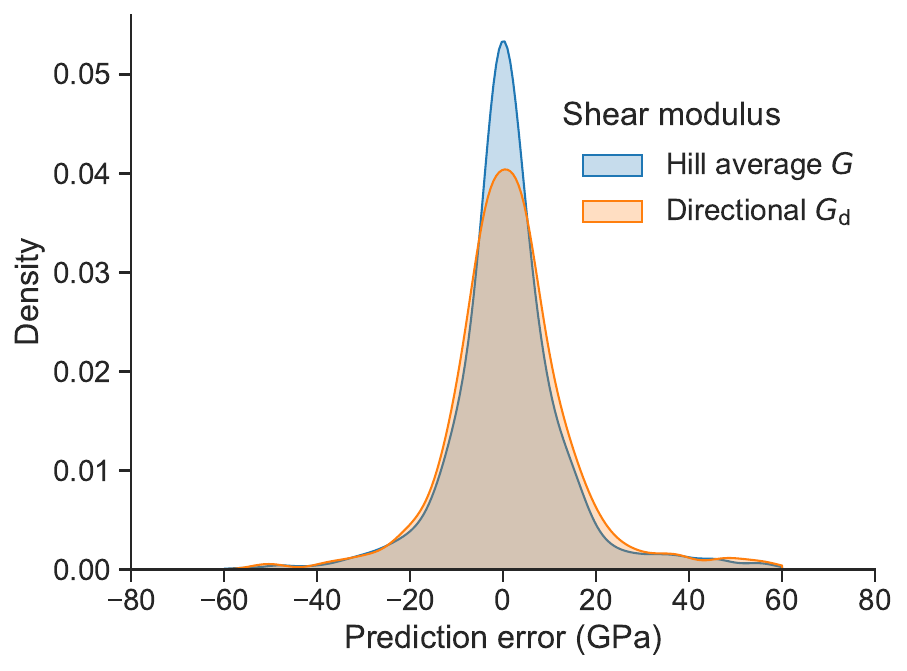}
    \caption{Distribution of the prediction error for shear modulus.
    The directional shear modulus can be computed as $G_\text{d}(\bm n, \bm m) = (n_im_jn_km_lS_{ijkl})^{-1}$, where $\bm n$ and $\bm m$ are two direction unit vectors, and $\bm S$ is the compliance tensor \cite{ran2023velas}.
    The data of $G_\text{d}$ is obtained by sampling in a way similar to $E_\text{d}$ discussed in the main text.
    The prediction error is the difference between \net prediction and DFT reference. 
    }
\end{figure}

\begin{figure}[H]
    \centering
    \includegraphics[width=1\columnwidth]{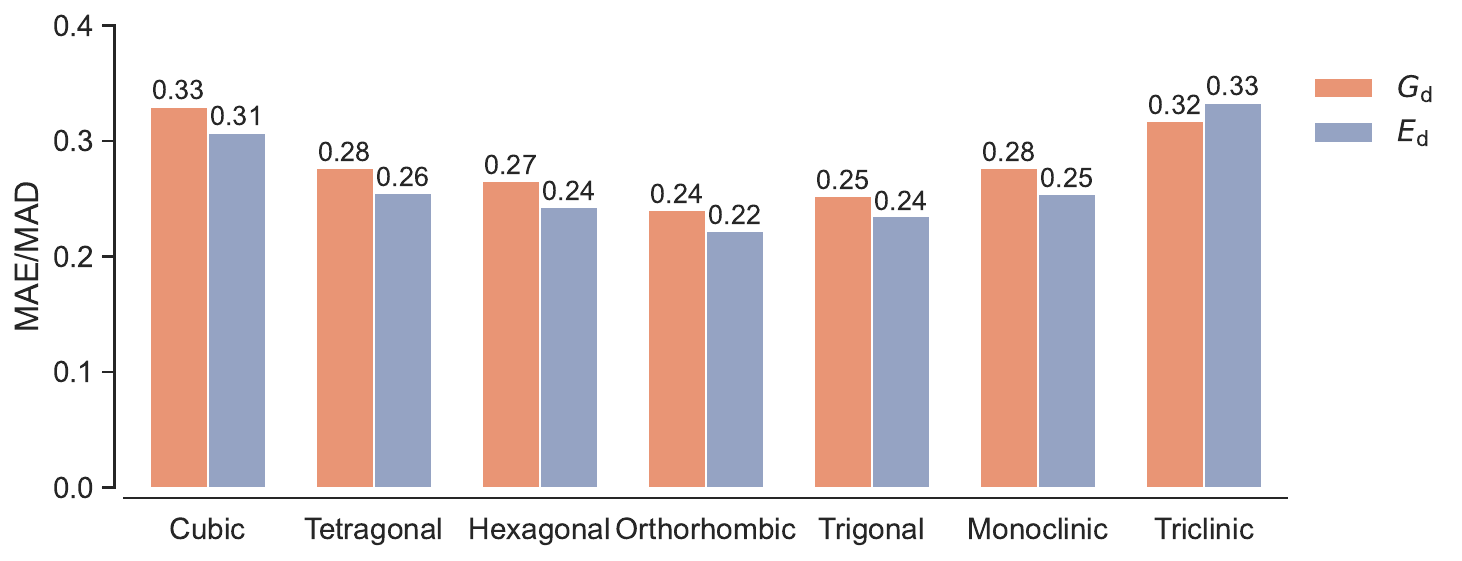}
    \caption{Scaled error in directional shear modulus $G_\text{d}$ and Young's modulus $E_\text{d}$.
    The bulk modulus has no directional dependency, and thus no such plot is presented.
    }
\end{figure}

\section*{Materials Screening}

\begin{figure}[H]
    \centering
    \includegraphics[width=0.95\columnwidth]{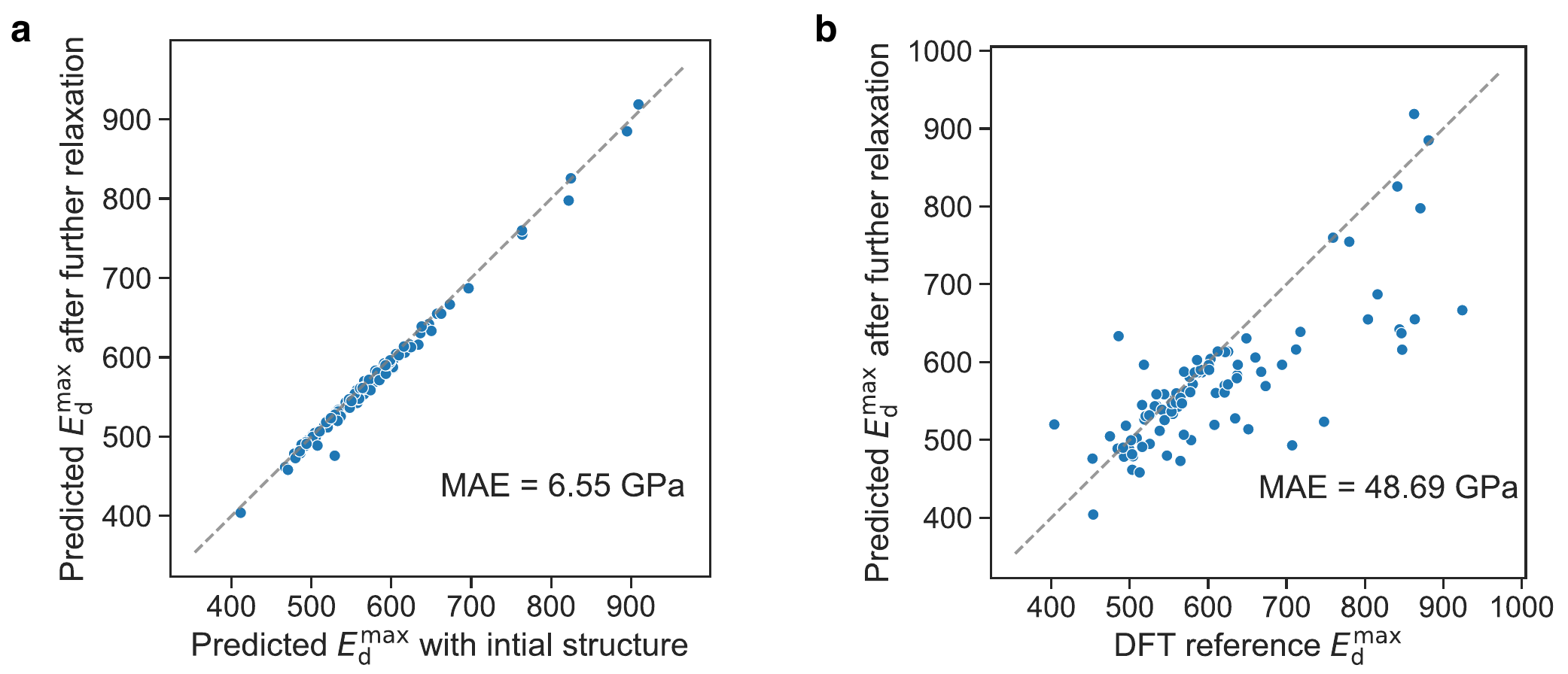}
    \caption{Maximum directional Young's modulus $E_\text{d}^\text{max}$ obtained from DFT and the \net model.
    Each material has two \net predictions, one using the crystal structure directly queried from the Materials Project (``Predicted $E_\text{d}^\text{max}$ with initial structure''), and the other using the crystal structure with further geometry optimization (``Predicted $E_\text{d}^\text{max}$ after further relaxation''). 
    The latter has tighter geometry optimization criteria. 
    The DFT reference $E_\text{d}^\text{max}$ is obtained using the latter further optimized geometry.
    }
    \label{fig:max:E:parity}
\end{figure}

\fref{fig:max:E:parity} shows the $E_\text{d}^\text{max}$ for the 100 new crystals. 
The MAE between predicted $E_\text{d}^\text{max}$ with initial structure and predicted $E_\text{d}^\text{max}$ with further relaxed structure is 6.55~GPa. 
It is much smaller than the MAE (22.36~GPa) between \net prediction and DFT reference for the test set. 
This demonstrates the robustness of \net with respect to the structure of the input crystal as discussed in the main text.
As shown in \fref{fig:max:E:parity}~b, if we consider the 100 new crystals instead of the test set, the MAE between \net prediction and DFT is much higher, with a value of 48.69~GPa. 
This is expected, since, for the 100 new crystals, we are probing extreme values at the edge of the training data distribution, while the test set follows the same distribution of the training data.
This signifies the importance of further confirmation with more accurate computation (DFT in this case) and even experiments once the search space has been narrowed down via the screening using the model.

\begin{table}[H]
\small
\caption{Polymorphs of elemental cubic metal with $E_\text{d}^\text{max}$ along $\langle 100 \rangle$ directions and $E_\text{d}^\text{min}$ along $\langle 111 \rangle$ directions. 
$\Delta S = S_{1111} - S_{1122} - 2S_{2323}$.
Among the crystal structures with the same composition, the one having the lowest energy is called the ground-state polymorph and is stable with respect to phase transition into other structures \cite{bartel2022review}.
The crystal structures and the elasticity tensors of these metals are provided as well. See Data Availability in the main text.
}
\centering
\begin{tabular}{@{\extracolsep{5pt}}cccccc}
\hline
Materials Project ID   & Formula   & $\Delta S_\text{DFT}$  & $\Delta S_\text{MatTen}$   & Experimentally observed  & Ground-state polymorph\\
\hline
mp-129    & Mo    & -0.00150   & -0.00191    & Yes   & Yes  \\
mp-146    & V    & -0.00994   & -0.01006    & Yes  & Yes \\
mp-17    & Cr    & -0.00267   & -0.00253    & Yes  & No \\
mp-90    & Cr    & -0.00369   & 0.00055    & Yes   & Yes\\
mp-91    & W    & -0.00056   & -0.00042  & Yes  & Yes \\
mp-11334    & W    & -0.00285   & -0.00331    & No   & No \\
mp-35    & Mn    & -0.00222   & -0.00277    & Yes  & Yes \\
mp-1186040    & Na    & -0.15435   & -0.32852    & No  & No \\
mp-1184808    & K    & -0.34397   & -0.12334    & No  & No \\
mp-949029    & Cs    & -0.53668   & -4.79427    & No  & No \\
mp-1239193    & Rh    & -0.03371   & -0.06590    & No  & No \\
mp-1187790    & Tl    & -0.09708   & -0.05340    & No  & No \\

\hline
\end{tabular}
\end{table}

\section*{Model hyperparameters}
\begin{table}[H]
\renewcommand{\arraystretch}{1.2} 
\small
\caption{Hyperparameter values obtained by grid search. ``fixed'' indicates no search, and the value is obtained based on previous work \cite{thomas2018tensor, batzner2022e3}.
Full set of the optimal hyperparameters is available in the ``pretrained
/20230627'' directory of the GitHub repo at: \url{https://github.com/wengroup/matten}.
}
\centering
\begin{tabular}{p{5cm}p{9cm}l}
\hline
Value   & Hyperparameter   & Searched values \\
\hline
 5\AA  &  cutoff radius to construct crystal graph, $r_\text{cut}$  & 4, 5, 6\\
\arrayrulecolor{gray} 
\hline
  16   & size of one-hot embedding vector for atomic species, $c$  & fixed \\
\hline
  8    & number of radial basis functions, $n$   & fixed\\
\hline
  3    & number of interaction blocks  & 2, 3, 4, 5\\
\hline
     32x0o+32x0e+16x1o+16x1e +4x2o+4x2e+2x3o+2x3e+2x4e   & irreducible representation of atom features in interaction blocks  & fixed \\
\hline

     0e+1o+2e+3o+4e   & irreducible representation of unit bond vector & fixed \\
\hline
     2   & number of MLP layers for embedding bond length as in $R_c$  & 2, 3, 4\\
\hline
     32  & number of nodes in the MLP for embedding bond length as in $R_c$  & 32, 64 \\
\hline
\arrayrulecolor{black} 
\hline
\end{tabular}
\end{table}

\newpage
\bibliographystyle{unsrtnat}